\documentclass[journal]{IEEEtran}
\usepackage{graphicx}
\usepackage{color}
\usepackage{placeins}
\usepackage{float}
\usepackage{tabularx,colortbl}
\usepackage{amssymb}
\usepackage{amsthm}
\usepackage{cite}
\usepackage{amsmath}
\usepackage{makecell}
\usepackage{subfigure}

\theoremstyle{plain}
\newtheorem{thm}{Theorem}

\newtheorem{coro}{Corollary}

\theoremstyle{plain}
\newtheorem{rem}{Remark}

\IEEEoverridecommandlockouts

\begin{document}


\title{Optimal Bilinear Equalizer for Cell-Free Massive MIMO Systems over Correlated Rician Channels 
\thanks{Z. Wang, J. Zhang, and B. Ai are with the State Key Laboratory of Advanced Rail Autonomous Operation, and also with the School of Electronics and Information Engineering, Beijing Jiaotong University, Beijing 100044, P. R. (e-mail: \{zhewang\_77, jiayizhang, boai\}@bjtu.edu.cn);}
\thanks{Emil Bj{\"o}rnson is with the Department of Computer Science, KTH Royal Institute of Technology, 114 28 Stockholm, Sweden (e-mail: emilbjo@kth.se);}
\thanks{D. Niyato is with the College of Computing \& Data Science, Nanyang Technological University, Singapore 639798 (e-mail: dniyato@ntu.edu.sg);}}
\author{Zhe Wang, Jiayi Zhang,~\IEEEmembership{Senior Member,~IEEE,} Emil Bj{\"o}rnson,~\IEEEmembership{Fellow,~IEEE}, \\ Dusit Niyato,~\IEEEmembership{Fellow,~IEEE}, Bo Ai,~\IEEEmembership{Fellow,~IEEE}\vspace*{-0.6cm}}
\maketitle

\begin{abstract}
In this paper, we explore the low-complexity optimal bilinear equalizer (OBE) combining scheme design for cell-free massive multiple-input multiple-output networks with spatially correlated Rician fading channels. We provide a spectral efficiency (SE) performance analysis framework for both the centralized and distributed processing schemes with bilinear equalizer (BE)-structure combining schemes applied. The BE-structured combining is a set of schemes that are constructed by the multiplications of channel statistics-based BE matrices and instantaneous channel estimates. Notably, we derive closed-form achievable SE expressions for centralized and distributed BE-structured combining schemes. We propose one centralized and two distributed OBE schemes: Centralized OBE (C-OBE), Distributed OBE based on Global channel statistics (DG-OBE), and Distributed OBE based on Local channel statistics (DL-OBE), which maximize their respective SE expressions. OBE matrices in these schemes are tailored based on varying levels of channel statistics. Notably, we obtain new and insightful closed-form results for the C-OBE, DG-OBE, and DL-OBE combining schemes. Numerical results demonstrate that the proposed OBE schemes can achieve excellent SE, even in scenarios with severe pilot contamination.

\end{abstract}

\IEEEpeerreviewmaketitle

\section{Introduction}
To achieve demanding requirements for wireless communication networks, multiple-input multiple-output (MIMO) technology has been regarded as the promising enabler since fourth-generation (4G) wireless communication networks \cite{biglieri2007mimo,5595728,1197843}. Notably, MIMO technology has gathered a big success in the fifth-generation (5G) wireless communication networks since it can deliver higher spectral efficiency in bands that feature advantageous propagation conditions \cite{6824752,8187178}. However, with the further development of wireless communication, sixth-generation (6G) networks are expected to meet substantially higher performance requirements \cite{you2021towards,8869705,9170653,Qiao-2024-Resource} and support a diverse array of emerging application scenarios \cite{pan2023resource,10670196,wang2022noma,ouyang2022performance,10679152}. To fulfill the anticipated performance enhancements, next-generation MIMO technologies have garnered significant research attention,  represented by cell-free massive MIMO (CF mMIMO) \cite{7827017,[162],buzzi2019user}, reconfigurable intelligent surfaces (RIS) \cite{8811733,9999288,enyusurvey}, and extremely large-scale MIMO (XL-MIMO) \cite{ZheSurvey,[32],9903389}. Among these technologies, CF mMIMO attracted considerable research interest due to its innovative network topology and its ability to provide uniform coverage.

The basic idea of CF mMIMO is to deploy a large number of access points (APs) in the coverage area, where all APs are connected to one or a few central processing units (CPUs) via fronthaul links \cite{interdonato2019ubiquitous,ngo2017total}. Relying on the coherent joint transmission and reception among APs and the support from the CPU, the CF mMIMO network delivers services to user equipment (UEs) through various processing methods, each differing in performance and complexity \cite{[162],04962}. To fully exploit the potential performance of CF mMIMO networks, the design of beamforming strategies is crucial. As demonstrated in \cite{[162]}, among various processing schemes, centralized processing with the minimum mean-square error (MMSE) combining 
can maximize the spectral efficiency (SE). This method is regarded as the benchmark for assessing the maximum achievable performance. To fully utilize the CF mMIMO network's innovative topology, it is interesting to explore the distributed processing scheme, where some processing tasks can be implemented at each AP locally. However, it is intrinsically hard to find anything optimal in the distributed case. A two-layer decoding scheme is advocated \cite{[162],04962}, where the local MMSE (L-MMSE) combining is applied at each AP as the first layer decoding, while the large-scale fading decoding (LSFD) scheme is implemented at the CPU as the second layer decoding. 

Although the centralized MMSE (C-MMSE) combining achieves excellent performance, it necessitates matrix inversion for high-dimensional matrices in every coherence block, which is computationally demanding. To balance the achievable performance and computational complexity, numerous alternative beamforming schemes have been developed. The authors in \cite{neumann2018bilinear} proposed a low-complexity alternative scheme, called optimal bilinear equalizer (OBE). In this scheme, the matrix inversion component in the C-MMSE combining is replaced with a matrix based on channel statistics, which eliminates the need for high-dimensional matrix inversion in every channel coherence block. It was observed that the average SE performance for the worst user under the OBE scheme could be approaching to that of the C-MMSE combining. However, \cite{neumann2018bilinear} considered the cellular mMIMO scenario. Based on the fundamentals in \cite{neumann2018bilinear}, the authors \cite{polegre2021pilot} studied a scheme called generalized maximum ratio (GMR) and showed that it performed well in scenarios with severe pilot contamination and could achieve the excellent SE performance in both the distributed and centralized configurations. However, the study in \cite{polegre2021pilot} considered CF mMIMO networks with single-antenna APs over Rayleigh fading channels. More generalized scenarios are also of interest to be further studied. On the one hand, APs in CF mMIMO networks will likely be equipped with multiple antennas in practice, resulting in spatially correlated characteristics. On the other hand, given the short distances between APs and UEs, the line-of-sight (LoS) path often becomes the dominant component of channels in practical CF mMIMO networks. Thus, CF mMIMO networks with multi-antenna APs over the spatially correlated Rician fading channels are advocated \cite{wang2020uplink}. Previous research \cite{wang2020uplink,liu2023cell} indicates that the inclusion of the LoS component and channel spatial correlation introduces significant challenges in performance evaluation and beamforming design, primarily due to complex matrix computations.

Based on the above motivations, we develop novel centralized and distributed OBE combining designs for CF mMIMO networks over spatially correlated Rician fading channels. The major contributions are listed as follows.   
\begin{itemize}
\item We develop an SE performance analysis framework for both the centralized and distributed processing schemes over spatially correlated Rician channels with bilinear equalize (BE)-structured combining schemes. Closed-form expressions of the achievable SE for both the centralized and distributed processing schemes are derived. 
\item For the centralized processing scheme, we study the OBE combining scheme, called centralized OBE (C-OBE), which can maximize the SE among all centralized BE-structured combining schemes. Furthermore, we present a novel closed-form computation of the C-OBE combining scheme.
\item Two distributed OBE schemes are proposed. The first, called Distributed OBE based on Global channel statistics (DG-OBE), utilizes the global channel statistics and local instantaneous channel estimates at each AP. The second one, called Distributed OBE based on Local channel statistics (DL-OBE), relies solely on the local channel statistics and estimates at each AP. We derive closed-form results for both the DG-OBE and DL-OBE schemes.
\end{itemize}

\textbf{\emph{Notation}}: Let boldface lowercase letters $\mathbf{x}$ and boldface uppercase letters $\mathbf{X}$ represent column vectors and matrices, respectively. We define $\mathbf{x}\sim \mathcal{N} _{\mathbb{C}}\left( \mathbf{0},\mathbf{R} \right)$ as a circularly symmetric complex Gaussian distribution vector with the correlation matrix $\mathbf{R}$. We denote $\left( \cdot \right) ^H$, $\left( \cdot \right) ^T$,  $\mathbb{E} \left\{ \cdot \right\}$, $\mathrm{tr}( \cdot )$, and $\mathrm{vec}\left( \cdot  \right)$  as the conjugate transpose, transpose, expectation operator, trace operator and vectorization operator, respectively. $\mathbf{I}_{N}$ represents the $N\times N$ identity matrix. $\otimes$ denotes the Kronecker product. $\left\| \cdot \right\|$ is the Euclidean norm and $\left| \cdot \right|$ denotes the absolute value of a number. Let $\mathrm{diag}\left( \mathbf{X}_1,\cdots ,\mathbf{X}_n \right)$ denote a block-diagonal matrix with the square matrices $\mathbf{X}_1,\cdots ,\mathbf{X}_n $ on the diagonal.

\vspace*{-0.3cm}

\section{Fundamentals of CF mMIMO Networks}\label{fundamentals}
In this paper, we study a CF mMIMO network, where $M$ APs with $N$ antennas each serve $K$ single-antenna UEs. All APs, which are connected via fronthaul to a CPU, and UEs are arbitrarily located in a wide coverage region. We define the channel response between AP $m$ and UE $k$ as $\mathbf{g}_{mk}\in \mathbb{C} ^N$, which remains constant in each time-frequency coherence block with $\tau_c$ channel uses \cite{8187178}. Note that the uplink pilot transmission and uplink data transmission are investigated. Thus, in each coherence time-frequency block with $\tau_c$ channel uses, $\tau_p$ and $\tau_c-\tau_p$ channel uses are reserved for pilots and data, respectively. 

\subsection{Channel Model}\label{channel}
We consider the spatially correlated Rician fading channel model, where both the LoS component and non-line-of-sight (NLoS) component are modelled. The channel between AP $m$ and UE $k$ can be modelled as
$
\mathbf{g}_{mk}=\overline{\mathbf{g}}_{mk}e^{j\theta _{mk}}+\check{\mathbf{g}}_{mk},
$
where $\overline{\mathbf{g}}_{mk}\in \mathbb{C} ^N$ is the semi-deterministic LoS component with $\theta _{mk}\sim \mathcal{U} [ -\pi ,\pi ]$ being the random phase-shift between AP $m$ and UE $k$ as in \cite{wang2020uplink} and $\check{\mathbf{g}}_{mk}\sim \mathcal{N} _{\mathbb{C}}\left( \mathbf{0},\check{\mathbf{R}}_{mk} \right)$ is the stochastic NLoS component. $\check{\mathbf{R}}_{mk}\in \mathbb{C} ^{N\times N}$ denotes the spatial correlation matrix with $\beta _{mk}^{\mathrm{NLoS}}={\mathrm{tr}( \check{\mathbf{R}}_{mk} )}/{N}$ being the large-scale fading coefficient for the NLoS component between AP $m$ and UE $k$. Meanwhile, we assume that $\check{\mathbf{g}}_{mk}$ for every AP-UE pair are independent and channel realizations of $\check{\mathbf{g}}_{mk}$ in different coherence blocks are independent and identically distributed.
\vspace*{-0.4cm}
\subsection{Channel Estimation}\label{CE}
We use a set of $\tau _p$ mutually orthogonal pilot signals, $\boldsymbol{\phi} _1,\dots ,\boldsymbol{\phi} _{\tau _p}$, during the channel estimation with $\left\| \boldsymbol{\phi} _t \right\| ^2=\tau _p$. Let $t_k\in\{1,\dots,\tau _p\}$ denote the index of the pilot signal sent by UE $k$. 
Thus, we denote the pilot signal sent by UE $k$ as $\boldsymbol{\phi} _{t_{k}}$. Due to the limited coherence block size, more than one UE would use the same pilot signal, that is $K>\tau _p$. We denote the subset of UEs, applying the same pilot signal as UE $k$, as $\mathcal{P} _k$ including UE $k$. All UEs transmit their respective allocated pilot signals and the received pilot signal at AP $m$ can be denoted as $\mathbf{Y}_{m}^{p}=\sum_{k=1}^K{\sqrt{p_k}\mathbf{g}_{mk}\boldsymbol{\phi} _{t_{k}}^{T}}+\mathbf{N}_{m}^{p}\in \mathbb{C} ^{N\times \tau _p}$,  where 
$p_k$ is the transmit power of UE $k$ and $\mathbf{N}_{m}^{p}\in \mathbb{C} ^{N\times \tau _p}$ is the additive noise with independent $\mathcal{N} _{\mathbb{C}}\left( 0,\sigma ^2\right)$ entries with $\sigma ^2$ being the noise power. Based on the fundamentals in \cite{wang2020uplink,8187178}, we derive the phase-aware MMSE estimate\footnote{Note that the phase-shifts vary through each coherence block. More specifically, the phase-shifts are constant over the frequency domain but vary randomly over the time domain at the same pace as the small-scale fading. Thus, each AP can be assumed to be aware to these time-varying phase-shifts.} of $\mathbf{g}_{mk}$ as
\begin{equation}\label{MMSE_CE}
\widehat{\mathbf{g}}_{mk}=\overline{\mathbf{g}}_{mk}e^{j\theta _{mk}}+\sqrt{p_k}\check{\mathbf{R}}_{mk}\mathbf{\Psi }_{mk}^{-1}\left( \mathbf{y}_{mk}^{p}-\overline{\mathbf{y}}_{mk}^{p} \right), 
\end{equation}
where $\mathbf{y}_{mk}^{p}=\mathbf{Y}_{m}^{p}\boldsymbol{\phi} _{t_k}^{*}=\sum_{l\in \mathcal{P} _k}{\sqrt{p_l}\tau _p\mathbf{g}_{ml}}+\mathbf{n}_{mk}^{p}$, $\overline{\mathbf{y}}_{mk}^{p}=\sum_{l\in \mathcal{P} _k}{\sqrt{p_l}\tau _p\overline{\mathbf{g}}_{ml}}e^{j\theta _{ml}}$, $\mathbf{n}_{mk}^{p}=\mathbf{N}_{m}^{p}\boldsymbol{\phi} _{k}^{*}\sim \mathcal{N} _{\mathbb{C}}( \mathbf{0},\tau _p\sigma ^2\mathbf{I}_N ) $, and
$
\mathbf{\Psi }_{mk}={\mathbb{E} \{ ( \mathbf{y}_{mk}^{p}-\overline{\mathbf{y}}_{mk}^{p} ) \left( \mathbf{y}_{mk}^{p}-\overline{\mathbf{y}}_{mk}^{p} \right) ^H \}}/{\tau _p}=\sum_{l\in \mathcal{P} _k}{p_l\tau _p\check{\mathbf{R}}_{ml}}+\sigma ^2\mathbf{I}_N.
$
The estimation error $\tilde{\mathbf{g}}_{mk}=\mathbf{g}_{mk}-\widehat{\mathbf{g}}_{mk} $ and estimate $\widehat{\mathbf{g}}_{mk}$ are independent random variables, where $\mathbb{E} \{ \widehat{\mathbf{g}}_{mk} |\theta _{mk} \} =\overline{\mathbf{g}}_{mk}e^{j\theta _{mk}}$, $\mathrm{Cov}\{ \widehat{\mathbf{g}}_{mk} |\theta _{mk} \} =\widehat{\mathbf{R}}_{mk}=p_k\tau _p\check{\mathbf{R}}_{mk}\mathbf{\Psi }_{mk}^{-1}\check{\mathbf{R}}_{mk}$, $\mathbb{E} \{ \tilde{\mathbf{g}}_{mk} \} =\mathbf{0}$, and $\mathrm{Cov} \{ \tilde{\mathbf{g}}_{mk} \} =\mathbf{C}_{mk}=\check{\mathbf{R}}_{mk}-p_k\tau _p\check{\mathbf{R}}_{mk}\mathbf{\Psi }_{mk}^{-1}\check{\mathbf{R}}_{mk}$, respectively.

\begin{rem}\label{Diff_CE}
When the phase-shifts are not known, alternative channel estimators, such as the linear MMSE (LMMSE) channel estimator $\widehat{\mathbf{g}}_{mk}=\sqrt{p_k}\check{\mathbf{R}}_{mk}^{\prime}\mathbf{\Psi }_{mk}^{\prime,-1}\mathbf{y}_{mk}^{p}$ \cite{wang2020uplink} and least-square (LS) channel estimator $\widehat{\mathbf{g}}_{mk}=1/\sqrt{p_k\tau _{p}^{2}}\mathbf{y}_{mk}^{p}$ \cite{ozdogan2019massive}, can be applied, where $\check{\mathbf{R}}_{mk}^{\prime}=\check{\mathbf{R}}_{mk}+\overline{\mathbf{g}}_{mk}\overline{\mathbf{g}}_{mk}^{H}$ and $\mathbf{\Psi }_{mk}^{\prime}=\sum_{l\in \mathcal{P} _k}{p_l\tau _p\check{\mathbf{R}}_{ml}^{\prime}}+\sigma ^2\mathbf{I}_N$.
\end{rem}

\begin{figure*}[t]
{{\begin{align}\tag{8}\label{Centrlized_closed}
\overline{\mathrm{SINR}}_{k}^{\mathrm{c},\mathrm{UatF}}=\frac{p_k\left| \mathrm{tr}\left( \mathbf{W}_{k}^{H}\overline{\mathbf{R}}_k \right) \right|^2}{\sum_{l=1}^K{p_l\mu _{kl}}+\sum_{l\in \mathcal{P} _k }{p_l\varepsilon _{kl}}+p_k\omega _k-p_k\left| \mathrm{tr}\left( \mathbf{W}_{k}^{H}\overline{\mathbf{R}}_k \right) \right|^2+\sigma ^2\mathrm{tr}\left( \mathbf{W}_k\overline{\mathbf{R}}_k\mathbf{W}_{k}^{H} \right)}
\end{align}}
\hrulefill
\vspace*{-0.6cm}
}\end{figure*}

\begin{figure*}[t]
{{\begin{align}\tag{10}\label{LoS_Closed}
&\omega _k=\!p_k\tau _p\mathrm{tr(}\mathbf{W}_{k}^{H}\check{\mathbf{R}}_k\mathbf{\Psi }_{k}^{-1}\check{\mathbf{R}}_k)\mathrm{tr(}\mathbf{W}_k\overline{\mathbf{G}}_{kk})+p_k\tau _p\mathrm{tr(}\mathbf{W}_{k}^{H}\overline{\mathbf{G}}_{kk})\mathrm{tr(}\mathbf{W}_k\check{\mathbf{R}}_k\mathbf{\Psi }_{k}^{-1}\check{\mathbf{R}}_k)
-\!\mathrm{tr(}\mathbf{W}_{k}^{H}\overline{\mathbf{G}}_{kk}\mathbf{W}_k\overline{\mathbf{G}}_{kk})\\\notag
&+\sum\nolimits_{m_1=1}^M{\sum\nolimits_{m_2=1,m_1\ne m_2}^M{[\mathrm{tr}( \mathbf{W}_{k,m_1m_1}^{H}\overline{\mathbf{G}}_{m_1m_2kk}\mathbf{W}_{k,m_2m_2}\overline{\mathbf{G}}_{m_1m_2kk}^{H} )}}+\mathrm{tr}( \mathbf{W}_{k,m_2m_1}^{H}\overline{\mathbf{G}}_{m_2kk}\mathbf{W}_{k,m_2m_1}\overline{\mathbf{G}}_{m_1kk} )]\\\notag
&+\sum\nolimits_{m_1=1}^M{\mathrm{tr}( \mathbf{W}_{k,m_1m_1}^{H}\overline{\mathbf{G}}_{m_1kk}\mathbf{W}_{k,m_1m_1}\overline{\mathbf{G}}_{m_1kk} )}
\end{align}}
\hrulefill
\vspace*{-0.5cm}
}\end{figure*}

\vspace*{-0.3cm}
\subsection{Data Transmission}\label{DT}
In the data transmission phase, all UEs transmit $\tau_c-\tau_p$ data symbols in each coherence block to all APs. The received data signal at AP $m$ $\mathbf{y}_m\in \mathbb{C} ^N$ can be denoted as
\begin{equation}\label{Data}
\mathbf{y}_m=\sum_{k=1}^K{\mathbf{g}_{mk}x_k}+\mathbf{n}_m,
\end{equation}
where $x_k\sim \mathcal{N} _{\mathbb{C}}( 0,p_k) $ is the data symbol transmitted by UE $k$ and $\mathbf{n}_m\sim \mathcal{N} _{\mathbb{C}}( \mathbf{0},\sigma ^2\mathbf{I}_N ) $ is the additive noise at AP $m$.

\section{Spectral Efficiency Analysis}\label{SE_Sec}
Based on the received data signal \eqref{Data} at all APs, in this section, we study the fully centralized processing scheme and the distributed processing scheme with the LSFD strategy. 
\vspace{-0.4cm}
\subsection{Centralized Processing}\label{centralized}
For the fully centralized processing scheme, all APs send all received pilot and data signals to the CPU via frounthaul links. The channel estimation and data detection are implemented at the CPU. Note that the channel for each UE $k$ can be constructed as a collective form as 
\begin{equation}
\mathbf{g}_k=\left[ \mathbf{g}_{1k}^{T},\dots ,\mathbf{g}_{Mk}^{T} \right] ^T=\overline{\mathbf{g}}_k+\check{\mathbf{g}}_k\in \mathbb{C} ^{MN\times 1}
\end{equation} 
where $\overline{\mathbf{g}}_k=[ \overline{\mathbf{g}}_{1k}^{T}e^{j\theta _{1k}},\dots ,\overline{\mathbf{g}}_{Mk}^{T}e^{j\theta _{Mk}} ] ^T$ and $\check{\mathbf{g}}_k=[ \check{\mathbf{g}}_{1k}^{T},\dots ,\check{\mathbf{g}}_{Mk}^{T} ] ^T$ denote the collective LoS and NLoS components for UE $k$, respectively. Relying on the independent characteristics for channels of different AP-UE pairs, we have 
$\check{\mathbf{R}}_k=\mathbb{E} \{ \check{\mathbf{g}}_k\check{\mathbf{g}}_{k}^{H} \} =\mathrm{diag}( \check{\mathbf{R}}_{1k},\dots ,\check{\mathbf{R}}_{Mk} ) \in \mathbb{C} ^{MN\times MN}$. Based on the received pilot signals from all APs $\{ \mathbf{Y}_{m}^{p}:m=1,\dots ,M\} $, the CPU can obtain the MMSE estimate of $\mathbf{g}_k$ as 
$\widehat{\mathbf{g}}_k=\overline{\mathbf{g}}_k+\sqrt{p_k}\check{\mathbf{R}}_k\mathbf{\Psi }_{k}^{-1}( \mathbf{y}_{k}^{p}-\overline{\mathbf{y}}_{k}^{p} )$,
where $\mathbf{y}_{k}^{p}=[ \mathbf{y}_{1k}^{p,T},\dots ,\mathbf{y}_{Mk}^{p,T} ] ^T\in \mathbb{C} ^{MN}$, $\overline{\mathbf{y}}_{k}^{p}=[ \overline{\mathbf{y}}_{1k}^{p,T},\dots ,\overline{\mathbf{y}}_{Mk}^{p,T} ] \in \mathbb{C} ^{MN}$, and $
\mathbf{\Psi }_k={\mathbb{E} \{ ( \mathbf{y}_{k}^{p}-\overline{\mathbf{y}}_{k}^{p} ) ( \mathbf{y}_{k}^{p}-\overline{\mathbf{y}}_{k}^{p} ) ^H \}}/{\tau _p}=\mathrm{diag}( \mathbf{\Psi }_{1k},\dots ,\mathbf{\Psi }_{Mk} ) \in \mathbb{C} ^{MN\times MN}$. The estimation error $\tilde{\mathbf{g}}_k=\mathbf{g}_k-\widehat{\mathbf{g}}_k$ and channel estimate $\widehat{\mathbf{g}}_k$ are also independent random variables with $\mathbb{E} \{ \widehat{\mathbf{g}}_{k} |\theta _{k} \} =\overline{\mathbf{g}}_{k}$, $\mathrm{Cov}\{ \widehat{\mathbf{g}}_{k} |\theta _{k} \} =\widehat{\mathbf{R}}_{k}=p_k\tau _p\check{\mathbf{R}}_{k}\mathbf{\Psi }_{k}^{-1}\check{\mathbf{R}}_{k}=\mathrm{diag}( \widehat{\mathbf{R}}_{1k},\dots ,\widehat{\mathbf{R}}_{Mk} )$, $\mathbb{E} \{ \tilde{\mathbf{g}}_{k} \} =\mathbf{0}$, $\mathrm{Cov} \{ \tilde{\mathbf{g}}_{k} \} =\mathbf{C}_{k}=\check{\mathbf{R}}_{k}-p_k\tau _p\check{\mathbf{R}}_{k}\mathbf{\Psi }_{k}^{-1}\check{\mathbf{R}}_{k}=\mathrm{diag}( \mathbf{C}_{1k},\dots ,\mathbf{C }_{Mk} )$, and $\theta _k=\{ \theta _{mk}:m=1,\dots ,M \}$, respectively.

By collecting the received data signals from all APs, the received data signal at the CPU can be denoted as
$
\mathbf{y}=\sum_{k=1}^K{\mathbf{g}_kx_k}+\mathbf{n},
$
where $\mathbf{n}\sim \mathcal{N} _{\mathbb{C}}( \mathbf{0},\sigma ^2\mathbf{I}_{MN}) $ is the additive noise. Arbitrary receive combining schemes $\mathbf{v}_k\in \mathbb{C} ^{MN}$ can be designed at the  CPU based on the channel estimates $\{ \widehat{\mathbf{g}}_k:k=1,\dots ,K \} $ and channel statistics to decode the data symbol transmitted by UE $k$ as
\begin{equation}\label{centralized_decode}
\widehat{x}_k=\mathbf{v}_{k}^{H}\mathbf{y}=\mathbf{v}_{k}^{H}\mathbf{g}_kx_k+\sum_{l\ne k}^K{\mathbf{v}_{k}^{H}\mathbf{g}_lx_l}+\mathbf{v}_{k}^{H}\mathbf{n}.
\end{equation}
According to \eqref{centralized_decode}, the achievable SE for UE $k$ can be computed based on the use-and-then-forget (UatF) bound \cite{8187178} as \begin{equation}\label{SE_centrlized_UatF}
\mathrm{SE}_{k}^{\mathrm{c},\mathrm{UatF}}=\frac{\tau _c-\tau _p}{\tau _c}\log _2\left( 1+\mathrm{SINR}_{k}^{\mathrm{c},\mathrm{UatF}} \right), 
\end{equation}
where $\mathrm{SINR}_{k}^{\mathrm{c},\mathrm{UatF}}$ is
\begin{equation}\label{SINR_centrlized_UatF}
\begin{aligned}
&\mathrm{SINR}_{k}^{\mathrm{c},\mathrm{UatF}}=\\
&\frac{p_k| \mathbb{E} \{ \mathbf{v}_{k}^{H}\mathbf{g}_k \} |^2}{\sum\limits_{l=1}^K{p_l\mathbb{E} \{ | \mathbf{v}_{k}^{H}\mathbf{g}_l |^2 \}}-p_k| \mathbb{E} \{ \mathbf{v}_{k}^{H}\mathbf{g}_k \} |^2\!+\!\sigma ^2\mathbb{E} \{ \| \mathbf{v}_k \| ^2 \}}.
\end{aligned}
\end{equation}
The expectations are derived with respect to all sources of randomness. 

\begin{figure*}[t]
{{\begin{align}\tag{22}\label{SINR_OBE}
\mathrm{SINR}_{k}^{\mathrm{c},\mathrm{UatF}}=\frac{p_k|\mathbb{E} \{\widehat{\mathbf{g}}_{k}^{H}\mathbf{W}_{k}^{H}\mathbf{g}_k\}|^2}{\sum\limits_{l=1}^K{p_l\mathbb{E} \{|\widehat{\mathbf{g}}_{k}^{H}\mathbf{W}_{k}^{H}\mathbf{g}_l|^2\}}-p_k|\mathbb{E} \{\widehat{\mathbf{g}}_{k}^{H}\mathbf{W}_{k}^{H}\mathbf{g}_k\}|^2\!+\!\sigma ^2\mathbb{E} \{\parallel \mathbf{W}_k\widehat{\mathbf{g}}_k\parallel ^2\}}.
\end{align}}
\hrulefill
\vspace*{-0.6cm}
}\end{figure*}

\begin{figure*}[t]
{{\begin{align}\tag{24}\label{Optimal_centralized_obe}
\mathbf{w}_{k}^{*}\!\!=\!\!\left( \sum_{l=1}^K{p_l\mathbb{E} \{  ( \widehat{\mathbf{g}}_k\widehat{\mathbf{g}}_{k}^{H} ) ^T\otimes ( \mathbf{g}_l\mathbf{g}_{l}^{H} )  \}}-p_k\mathrm{vec}( \mathbb{E} \{ \mathbf{g}_k\widehat{\mathbf{g}}_{k}^{H} \} ) \mathrm{vec}( \mathbb{E} \{ \mathbf{g}_k\widehat{\mathbf{g}}_{k}^{H} \} ) ^H+( \mathbb{E} \{ \widehat{\mathbf{g}}_k\widehat{\mathbf{g}}_{k}^{H} \} ^T\otimes \mathbf{I}_{MN} ) \right) ^{-1}\mathrm{vec}( \mathbb{E} \{ \mathbf{g}_k\widehat{\mathbf{g}}_{k}^{H} \} )
\end{align}}
\hrulefill
\vspace*{-0.6cm}
}\end{figure*}

We study a generalized combining scheme as 
\begin{equation}\label{Centrlized_Wg}
\mathbf{v}_k=\mathbf{W}_k\widehat{\mathbf{g}}_k,
\end{equation}
where $\mathbf{W}_k\in \mathbb{C} ^{MN\times MN}$ is an arbitrary matrix determined only by channel statistics instead of channel estimates. Alternatively, the conventional C-MMSE combining as 
$\mathbf{v}_k=p_k( \sum_{l=1}^K{p_l\left( \widehat{\mathbf{g}}_l\widehat{\mathbf{g}}_{l}^{H}+\mathbf{C}_l \right)}+\sigma ^2\mathbf{I}_{MN} ) ^{-1}\widehat{\mathbf{g}}_k$ \cite{[162]} is also on the form of \eqref{Centrlized_Wg}, except that $\mathbf{W}_k$ in the C-MMSE combining depends on the instantaneous channel estimates. Note that $\mathbf{W}_k$ remains constant in a long period of time since the channel statistics change slowly in time \cite{8187178}. Thus, by applying the combining scheme in \eqref{Centrlized_Wg}, only one matrix-vector multiplication is required for the combining scheme design for each UE in a coherence block, avoiding computationally demanding instantaneous information-based matrix inversion.

When applying the combining scheme in \eqref{Centrlized_Wg}, we can compute \eqref{SE_centrlized_UatF} in novel closed-form as follows.
\vspace*{-0.3cm}
\begin{thm}\label{thm_centralized_closed}
With the combining scheme in \eqref{Centrlized_Wg}, the achievable SE for UE $k$ in \eqref{SE_centrlized_UatF} can be computed in closed-form as $\overline{\mathrm{SE}}_{k}^{\mathrm{c},\mathrm{UatF}}=\frac{\tau _c-\tau _p}{\tau _c}\log _2( 1+\overline{\mathrm{SINR}}_{k}^{\mathrm{c},\mathrm{UatF}} ) $ with $\overline{\mathrm{SINR}}_{k}^{\mathrm{c},\mathrm{UatF}}$ given in \eqref{Centrlized_closed} on the top of this page, where $\varepsilon _{kl}=p_kp_l\tau _{p}^{2}| \mathrm{tr}( \mathbf{W}_{k}^{H}\check{\mathbf{R}}_l\mathbf{\Psi }_{k}^{-1}\check{\mathbf{R}}_k ) |^2$,
\setcounter{equation}{8}
\begin{equation}\label{mu}
\begin{aligned}
\mu _{kl}&=\mathrm{tr}( \mathbf{W}_{k}^{H}\overline{\mathbf{G}}_{ll}\mathbf{W}_k\overline{\mathbf{G}}_{kk} ) +\mathrm{tr}( \mathbf{W}_{k}^{H}\check{\mathbf{R}}_l\mathbf{W}_k\overline{\mathbf{G}}_{kk} )\\
&+\mathrm{tr}( \mathbf{W}_{k}^{H}\overline{\mathbf{G}}_{ll}\mathbf{W}_k\widehat{\mathbf{R}}_k ) +\mathrm{tr}( \mathbf{W}_{k}^{H}\check{\mathbf{R}}_l\mathbf{W}_k\widehat{\mathbf{R}}_k ), 
\end{aligned}
\end{equation}
$\omega _k$ is given in \eqref{LoS_Closed} on the top of this page,
$\overline{\mathbf{G}}_{kk}=\mathrm{diag}[ \overline{\mathbf{G}}_{1kk},\dots ,\overline{\mathbf{G}}_{Mkk} ]\in \mathbb{C} ^{MN\times MN} $ with $\overline{\mathbf{G}}_{mkk}=\overline{\mathbf{g}}_{mk}\overline{\mathbf{g}}_{mk}^{H}\in \mathbb{C} ^{N\times N}$, $\overline{\mathbf{R}}_k=\overline{\mathbf{G}}_{kk}+p_k\tau _p\check{\mathbf{R}}_k\mathbf{\Psi }_{k}^{-1}\check{\mathbf{R}}_k \in \mathbb{C} ^{MN\times MN}$, $\overline{\mathbf{G}}_{m_1m_2kk}=\overline{\mathbf{g}}_{m_1k}\overline{\mathbf{g}}_{m_2k}^{H}$, and $\mathbf{W}_{k,m_1m_2}\in \mathbb{C} ^{N\times N}$ denotes the $(m_1,m_2)$-th sub-matrix of $\mathbf{W}_{k}$, respectively.
\end{thm}

\begin{IEEEproof}
The proof is provided in Appendix~\ref{app_cenralized_closed}.
\end{IEEEproof}

\begin{rem}
Note that closed-form results in Theorem~\ref{thm_centralized_closed} are applicable for any deterministic $\mathbf{W}_{k}$. When $\mathbf{W}_{k}=\mathbf{I}_{MN}$, the results in Theorem~\ref{thm_centralized_closed} simplify to those of centralized MR (C-MR) combining such as \cite[Corollary 2]{li2023spatially}. Moreover, the effects of the semi-deterministic LoS component can be clearly observed from \eqref{Centrlized_closed}, which are indicated by the $\overline{\mathbf{G}}_{kk}$-related terms. By letting $\overline{\mathbf{G}}_{kk}=\mathbf{0}$, we can derive the closed-form results for the Rayleigh fading channel model, which consists of only the NLoS component.
\end{rem}

\vspace*{-0.6cm}

\begin{figure*}[t]
{{\begin{align}\tag{26}\label{SINR_Distributed}
\widetilde{\mathrm{SINR}}_{k}^{\mathrm{LSFD}}=\frac{p_k| \sum_{m=1}^M{\mathbb{E} \{ \mathbf{v}_{mk}^{H}\mathbf{g}_{mk} \}} |^2}{\sum_{l=1}^K{p_l\mathbb{E} \{ | \sum_{m=1}^M{\mathbf{v}_{mk}^{H}\mathbf{g}_{ml}} |^2 \} -p_k| \sum_{m=1}^M{\mathbb{E} \{ \mathbf{v}_{mk}^{H}\mathbf{g}_{mk} \}} |^2+\sigma ^2\sum_{m=1}^M{\mathbb{E} \{ \| \mathbf{v}_{mk} \| ^2 \}}}}
\end{align}}
\hrulefill
\vspace*{-0.43cm}
}\end{figure*}

\subsection{Distributed Processing}\label{distributed}
\vspace*{-0.1cm}
In this subsection, we study the distributed processing implementation. Based on the received data signal as \eqref{Data}, each AP can first compute soft estimates of the data and then send them to the CPU for final decoding based on LSFD strategy \cite{[162]}.
Relying on the local combining vector designed for UE $k$ $\mathbf{v}_{mk}\in \mathbb{C} ^N$, AP $m$ can obtain its local estimate of $x_k$ as
\setcounter{equation}{10}
\begin{equation}\label{local_decode}
\begin{aligned}
\widehat{x}_{mk}\!=\!\mathbf{v}_{mk}^{H}\mathbf{y}_m\!=\!\mathbf{v}_{mk}^{H}\mathbf{g}_{mk}x_k\!\!+\!\!\sum_{l\ne k}^K{\mathbf{v}_{mk}^{H}\mathbf{g}_{ml}x_l}\!+\!\mathbf{v}_{mk}^{H}\mathbf{n}_m.
\end{aligned}
\end{equation}
One efficient combining scheme is the local MMSE (L-MMSE) combining as
\begin{equation}\label{local_MMSE}
\begin{aligned}
\mathbf{v}_{mk}=p_k\left( \sum_{l=1}^K{p_l\left( \widehat{\mathbf{g}}_{ml}\widehat{\mathbf{g}}_{ml}^{H}+\mathbf{C}_{ml} \right)}+\sigma ^2\mathbf{I}_N \right) ^{-1}\widehat{\mathbf{g}}_{mk},
\end{aligned}
\end{equation}
which can minimize the local conditional MSE, $\mathrm{MSE}_{mk}=\mathbb{E} \{  | x_k-\widehat{x}_{mk} |^2 |\widehat{\mathbf{g}}_{mk} \} $. In the distributed processing scheme, the L-MMSE combining is widely implemented at each AP due to its outstanding performance \cite{[162],liu2023cell}. However, the L-MMSE combining is designed with $N\times N$ matrix inversion at every coherence block, which is computationally demanding when each AP is equipped with a large number of antennas, which might happen in 6G networks with high frequencies. Inspired by the structure of \eqref{local_MMSE}, we study a generalized distributed combining scheme like the form of \eqref{Centrlized_Wg} as
\begin{equation}\label{local_generalized}
\begin{aligned}
\mathbf{v}_{mk}=\mathbf{W}_{mk}\widehat{\mathbf{g}}_{mk},
\end{aligned}
\end{equation}
where $\mathbf{W}_{mk}\in \mathbb{C} ^{N\times N}$ is an arbitrary matrix determined only by channel statistics.

After computing local data estimates as \eqref{local_decode}, all APs send them to the CPU to derive the final decoding by applying the LSFD weighting coefficients for UE $k$ $\{ a_{mk}:m=1,\dots ,M \} $ estimates as $\check{x}_k=\sum_{m=1}^M{a_{mk}^{*}\widehat{x}_{mk}}=( \sum_{m=1}^M{a_{mk}^{*}\mathbf{v}_{mk}^{H}\mathbf{g}_{mk}} ) x_k+\sum_{l\ne k}^K{\sum_{m=1}^M{a_{mk}^{*}\mathbf{v}_{mk}^{H}\mathbf{g}_{ml}x_l}}+\mathbf{n}_{k}^{\prime}$, where $\mathbf{n}_{k}^{\prime}=\sum_{m=1}^M{a_{mk}^{*}\mathbf{v}_{mk}^{H}\mathbf{n}_m}$. Note that only the decoding data estimates are sent to the CPU thus it does not know the instantaneous channel estimates. Thus, based on $\check{x}_k$, by applying the UatF capacity bound, we can derive the achievable SE expression for the distributed processing scheme as \cite{[162]}
\begin{equation}\label{SE_LSFD}
\mathrm{SE}_{k}^{\mathrm{LSFD}}=\frac{\tau _c-\tau _p}{\tau _c}\log _2(1+\mathrm{SINR}_{k}^{\mathrm{LSFD}}),
\end{equation}
with
\begin{equation}\label{SINR_LSFD}
\mathrm{SINR}_{k}^{\mathrm{LSFD}}\!\!=\!\!\frac{p_k\left| \mathbf{a}_{k}^{H}\mathbb{E} \left\{ \mathbf{b}_{kk} \right\} \right|^2}{\mathbf{a}_{k}^{H}\!( \sum\limits_{l=1}^K{p_l\mathbf{\Xi }_{kl}}\!-\!p_k\mathbb{E} \left\{ \mathbf{b}_{kk} \right\} \mathbb{E} \left\{ \mathbf{b}_{kk}^{H} \right\} +\sigma ^2\mathbf{D}_k ) \mathbf{a}_k},
\end{equation}
where $\mathbf{a}_k=[ a_{1k},\dots ,a_{Mk} ] \in \mathbb{C} ^M$, $\mathbf{b}_{kk}=[ \mathbf{v}_{1k}^{H}\mathbf{g}_{1k},\dots ,\mathbf{v}_{Mk}^{H}\mathbf{g}_{Mk} ] ^T\in \mathbb{C} ^M$, $\mathbf{D}_k=\mathrm{diag}[ \mathbb{E} \{ \| \mathbf{v}_{1k} \| ^2 \} ,\dots ,\mathbb{E} \{ \| \mathbf{v}_{Mk} \| ^2 \} ] \in \mathbb{C} ^{M\times M}$, and $\mathbf{\Xi }_{kl}\in \mathbb{C} ^{M\times M}$ with its $(n,m)$-th element being $[ \mathbf{\Xi }_{kl} ] _{nm}=\mathbb{E} \{( \mathbf{v}_{mk}^{H}\mathbf{g}_{ml} ) ^H( \mathbf{v}_{nk}^{H}\mathbf{g}_{nl} ) \}$. Moreover, Note that \eqref{SINR_LSFD} is a standard Rayleigh quotient with respect to $\mathbf{a}_k$. Thus, we can derive the optimal $\mathbf{a}_k$ to maximize \eqref{SINR_LSFD} as 
\begin{equation}\label{optimal_lsfd_eq}
\mathbf{a}_{k}^{*}\!\!=\!\!\left( \sum_{l=1}^K{p_l\mathbf{\Xi }_{kl}}-p_k\mathbb{E} \left\{ \mathbf{b}_{kk} \right\} \mathbb{E} \left\{ \mathbf{b}_{kk}^{H} \right\} +\sigma ^2\mathbf{D}_k \right) ^{-1}\mathbb{E} \left\{ \mathbf{b}_{kk} \right\} 
\end{equation}
with the maximum SINR value as $\mathrm{SINR}_{k}^{\mathrm{LSFD},*}=p_k\mathbb{E} \left\{ \mathbf{b}_{kk}^{H} \right\} \mathbf{a}_{k}^{*}$ \cite{[162]}.

We can compute \eqref{SE_LSFD} in novel closed-form when the combining scheme in \eqref{local_generalized} is applied as the following theorem.
\begin{thm}\label{thm_LSFD_closed}
When the combining scheme in \eqref{local_generalized} is applied, we can derive closed-form SE expressions for the LSFD processing scheme as
\begin{equation}\label{SE_LSFD_Closed}
\overline{\mathrm{SE}}_{k}^{\mathrm{LSFD}}=\frac{\tau _c-\tau _p}{\tau _c}\log _2(1+\overline{\mathrm{SINR}}_{k}^{\mathrm{LSFD}}), 
\end{equation}
with 
\begin{equation}\label{SINR_LSFD_Closed}
\overline{\mathrm{SINR}}_{k}^{\mathrm{LSFD}}=\frac{p_k| \mathbf{a}_{k}^{H}\overline{\mathbf{b}}_{kk} |^2}{\mathbf{a}_{k}^{H}( \sum_{l=1}^K{p_l\overline{\mathbf{\Xi }}_{kl}}-p_k\overline{\mathbf{b}}_{kk}\overline{\mathbf{b}}_{kk}^{H}+\sigma ^2\overline{\mathbf{D}}_k ) \mathbf{a}_k},
\end{equation}
where $\overline{\mathbf{b}}_{kk}=[ \mathrm{tr}( \mathbf{W}_{1k}^{H}\overline{\mathbf{R}}_{1k} ) ,\dots ,\mathrm{tr}( \mathbf{W}_{Mk}^{H}\overline{\mathbf{R}}_{Mk} ) ] ^T\in \mathbb{C} ^M$ and $\overline{\mathbf{D}}_k=\mathrm{diag}[ \mathrm{tr}( \mathbf{W}_{1k}\overline{\mathbf{R}}_{1k} \mathbf{W}_{1k}^{H}) ,\dots ,\mathrm{tr}( \mathbf{W}_{Mk}\overline{\mathbf{R}}_{Mk} \mathbf{W}_{Mk}^{H}) ]$ $ \in \mathbb{C} ^{M\times M}$, respectively. As for $\overline{\mathbf{\Xi }}_{kl}\in \mathbb{C} ^{M\times M}$, if $l\notin \mathcal{P} _k$, the $(n,m)$-th element of $\overline{\mathbf{\Xi }}_{kl}$ is $[ \overline{\mathbf{\Xi }}_{kl} ] _{nm}=\varphi _{mkl} $ for $m=n$ with
\begin{equation}\label{local_var}
\begin{aligned}
\varphi _{mkl}\!\!&=\!\!\mathrm{tr}( \mathbf{W}_{mk}^{H}\overline{\mathbf{G}}_{mll}\mathbf{W}_{mk}\overline{\mathbf{G}}_{mkk} ) \!+\!\mathrm{tr}( \mathbf{W}_{mk}^{H}\check{\mathbf{R}}_{ml}\mathbf{W}_{mk}\overline{\mathbf{G}}_{mkk} ) \\
&\!+\!\mathrm{tr}( \mathbf{W}_{mk}^{H}\overline{\mathbf{G}}_{mll}\mathbf{W}_{mk}\widehat{\mathbf{R}}_{mk} ) +\mathrm{tr}( \mathbf{W}_{mk}^{H}\check{\mathbf{R}}_{ml}\mathbf{W}_{mk}\widehat{\mathbf{R}}_{mk}), 
\end{aligned}
\end{equation}
and $[ \overline{\mathbf{\Xi }}_{kl} ] _{nm}=0 $ for $m\ne n$. If $l\in \mathcal{P} _k$, the $(n,m)$-th element of $\overline{\mathbf{\Xi }}_{kl}$ is  $[ \overline{\mathbf{\Xi }}_{kl} ] _{nm}=\varphi _{mkl}+\upsilon _{mkl}$ for $m=n$ with
\begin{equation}
\begin{aligned}
&\upsilon _{mkl}=p_kp_l\tau _{p}^{2}| \mathrm{tr}( \mathbf{W}_{mk}^{H}\check{\mathbf{R}}_{ml}\mathbf{\Psi }_{mk}^{-1}\check{\mathbf{R}}_{mk} ) |^2+\\
&\begin{cases}
	0, \,\, \mathrm{if} \,\,l\in \mathcal{P} _k\backslash \left\{ k \right\}\\
	p_k\tau _p\mathrm{tr}( \mathbf{W}_{mk}^{H}\overline{\mathbf{G}}_{mkk} ) \mathrm{tr}( \mathbf{W}_{mk}\check{\mathbf{R}}_{mk}\mathbf{\Psi }_{mk}^{-1}\check{\mathbf{R}}_{mk} )\\
	+p_k\tau _p\mathrm{tr}( \mathbf{W}_{mk}^{H}\check{\mathbf{R}}_{mk}\mathbf{\Psi }_{mk}^{-1}\check{\mathbf{R}}_{mk} ) \mathrm{tr}( \mathbf{W}_{mk}\overline{\mathbf{G}}_{mkk} ) ,\,\, \mathrm{if} \,\,l=k\\
\end{cases}
\end{aligned}
\end{equation}
and $[\overline{\mathbf{\Xi }}_{kl} ] _{nm}=\mathrm{tr}( \mathbf{W}_{mk}\mathbf{B}_{mkl} ) \mathrm{tr}( \mathbf{W}_{nk}^{H}\mathbf{B}_{nlk}) $ for $m\ne n$ with $\mathbf{B}_{mlk}=\sqrt{p_kp_l}\tau _p\check{\mathbf{R}}_{ml}\mathbf{\Psi }_{mk}^{-1}\check{\mathbf{R}}_{mk}$ for $l\in \mathcal{P} _k\backslash \{ k \}$ and $\mathbf{B}_{mkk}=\overline{\mathbf{G}}_{mkk}+\sqrt{p_kp_l}\tau _p\check{\mathbf{R}}_{mk}\mathbf{\Psi }_{mk}^{-1}\check{\mathbf{R}}_{ml}$ for $l=k$, and 
$\overline{\mathbf{R}}_{mk}=\overline{\mathbf{G}}_{mkk}+p_k\tau _p\check{\mathbf{R}}_{mk}\mathbf{\Psi }_{mk}^{-1}\check{\mathbf{R}}_{mk} $, respectively.

Moreover, we can also denote the optimal $\mathbf{a}_k$ and the maximum value of $\mathrm{SINR}_{k}^{\mathrm{LSFD}}$ in closed-form as 
\begin{equation}
\overline{\mathbf{a}}_{k}^{*}=\left( \sum_{l=1}^K{p_l\overline{\mathbf{\Xi }}_{kl}}-p_k\overline{\mathbf{b}}_{kk}\overline{\mathbf{b}}_{kk}^{H}+\sigma ^2\overline{\mathbf{D}}_k \right) ^{-1}\overline{\mathbf{b}}_{kk}
\end{equation} 
and 
$\overline{\mathrm{SINR}}_{k}^{\mathrm{LSFD},*}=p_k\overline{\mathbf{b}}_{kk}^{H}\overline{\mathbf{a}}_{k}^{*}$, respectively.
\end{thm}
\begin{IEEEproof}
The proof is provided in Appendix~\ref{closed-LSFD-app}.
\end{IEEEproof}

{\begin{rem}
Note that \eqref{SINR_LSFD} is derived based on the UatF capacity bound, which holds for arbitrary receive combining vectors $\mathbf{v}_{mk}$. Indeed, for each realization of the AP/UE locations, \eqref{SINR_LSFD} can be computed via the Monte-Carlo simulation methodology by considering many instantaneous channel realizations. When the OBE combining scheme $\mathbf{v}_{mk}=\mathbf{W}_{mk}\widehat{\mathbf{g}}_{mk}$ in \eqref{local_generalized} is applied, the expectations in \eqref{SINR_LSFD} can be computed in closed-form in \eqref{SINR_LSFD_Closed} based on the channel statistics.
\end{rem}

\begin{rem}
The derived closed-form results in Theorem~\ref{thm_LSFD_closed} are applicable for any configuration of $\mathbf{W}_{mk}$. When $\mathbf{W}_{mk}=\mathbf{I}_{N }$, derived results can reduce to those of L-MR combining such as \cite[Sec. III-A]{wang2020uplink} for the L-MR combining based LSFD processing scheme over the Rician fading channel.
\end{rem}

\vspace*{-0.5cm}

\section{Optimal Bilinear Equalizer Design}\label{sec_obe}
The combining schemes in \eqref{Centrlized_Wg} and \eqref{local_generalized} can be named as BE. This is because the decoding estimates in \eqref{centralized_decode} and \eqref{local_decode} are not only linear to the received data vectors $\mathbf{y}_m$ and $\mathbf{y}_{mk}$ but also linear to the channel estimates $\widehat{\mathbf{g}}_k$ and $\widehat{\mathbf{g}}_{mk}$, for the centralized and distributed processing schemes, respectively \cite{neumann2018bilinear}. In this section, we will optimally design $\mathbf{W}_{k}$ and $\mathbf{W}_{mk}$ to derive the OBE which can achieve the best achievable SE performance.
\vspace{-0.2cm}
\subsection{C-OBE Combining Scheme Design}\label{Sec-C-OBE}
For the centralized processing scheme, we optimally design the C-OBE, which can maximize the achievable SE for the centralized processing scheme based on the UatF capacity bound in Sec.~\ref{centralized}. When applying the BE combining scheme \eqref{Centrlized_Wg}, we can represent the UatF bound based SINR in \eqref{SINR_centrlized_UatF} in \eqref{SINR_OBE} on the top of the previous page.
Furthermore, we can derive the optimal $\mathbf{W}_{k}$, which can maximize \eqref{SINR_centrlized_UatF} as follows\footnote{In the following, we focus on the analysis of the SINR. The achievable SE can be easily derived from the SINR based on the above results. Notably, when the SINR is maximized, the achievable SE can also be maximized since $\tau_c$ and $\tau_p$ are constants in this paper.}.
\begin{coro}\label{C_OBE_UatF}
The UatF bound based SINR in \eqref{SINR_centrlized_UatF} can be maximized by $\mathbf{W}_{k}^{*}=\mathrm{vec}^{-1}( \mathbf{w}_{k}^{*} ) $ with the maximum value as
\setcounter{equation}{22}
\begin{equation}\label{C_SINR_Max}
\mathrm{SINR}_{k}^{\mathrm{c},\mathrm{UatF},*}=p_k\mathrm{vec}( \mathbb{E} \{ \mathbf{g}_k\widehat{\mathbf{g}}_{k}^{H} \} ) ^H\mathbf{w}_{k}^{*},
\end{equation}
where $\mathbf{w}_{k}^{*}\in \mathbb{C} ^{M^2N^2}$ is given in \eqref{Optimal_centralized_obe} on the top of the previous page.
\end{coro}
\begin{IEEEproof}
We provide the proof in Appendix~\ref{app_centralized_obe}.
\end{IEEEproof}

Based on Theorem~\ref{thm_centralized_closed}, we can compute $\mathbf{W}_{k}^{*}$ in novel closed-form as follows.
\begin{thm}\label{Centralized_OBE_closed}
The closed-form results of the OBE matrix $\mathbf{W}_{k}^{*}$ can be computed as $\overline{\mathbf{W}}_{k}^{*}=\mathrm{vec}^{-1}( \overline{\mathbf{w}}_{k}^{*} ) $, with
\setcounter{equation}{24}
\begin{equation}\label{Centralized_obe_vector}
\overline{\mathbf{w}}_{k}^{*}=\mathbf{\Gamma }_{k}^{-1}\overline{\mathbf{r}}_k,
\end{equation}
where $\mathbf{\Gamma }_k=\sum\nolimits_{l=1}^K{p_l[ ( \overline{\mathbf{G}}_{kk}^{T}\otimes \overline{\mathbf{G}}_{ll} ) +( \overline{\mathbf{G}}_{kk}^{T}\otimes \check{\mathbf{R}}_l ) } +( \widehat{\mathbf{R}}_{k}^{T}\otimes \overline{\mathbf{G}}_{ll} ) +( \widehat{\mathbf{R}}_{k}^{T}\otimes \check{\mathbf{R}}_l ) ]+\sum\nolimits_{l\in \mathcal{P} _k}{p_kp_{l}^{2}\tau _{p}^{2}\tilde{\mathbf{r}}_{lk}\tilde{\mathbf{r}}_{lk}^{H}}+p_{k}^{2}\tau _p\overline{\mathbf{g}}_{kk}\tilde{\mathbf{r}}_{kk}^{H}+p_{k}^{2}\tau _p\tilde{\mathbf{r}}_{kk}\overline{\mathbf{g}}_{kk}^{H}+p_k\overline{\mathbf{\Upsilon }}_k -p_k( \overline{\mathbf{G}}_{kk}^{T}\otimes \overline{\mathbf{G}}_{kk} ) -p_k\overline{\mathbf{r}}_k\overline{\mathbf{r}}_{k}^{H}+\sigma ^2( \overline{\mathbf{R}}_{k}^{T}\otimes \mathbf{I}_{MN} )\in \mathbb{C} ^{M^2N^2 \times M^2N^2}$ with $\tilde{\mathbf{r}}_{lk}=\mathrm{vec}( \check{\mathbf{R}}_l\mathbf{\Psi }_{k}^{-1}\check{\mathbf{R}}_k )$, $\overline{\mathbf{g}}_{kk}=\mathrm{vec}( \overline{\mathbf{G}}_{kk} ) $, and $\overline{\mathbf{r}}_k=\mathrm{vec}( \overline{\mathbf{R}}_k ) $. As for $\overline{\mathbf{\Upsilon }}_k\in\mathbb{C} ^{M^2N^2 \times M^2N^2}$, the $(a,b)$-th element of the $(c,d)$-th block matrix of it $\overline{\mathbf{\Upsilon }}_{k,cd}\in\mathbb{C} ^{MN\times MN}$ with $\{a,b,c,d\}=1,\dots ,MN$ is $[ \overline{\mathbf{\Upsilon }}_{k,cd} ] _{ab}=\overline{g}_{m_1k,n_1}\overline{g}_{m_2k,n_2}^{*}\overline{g}_{m_3k,n_3}\overline{g}_{m_4k,n_4}^{*}$ if $m_1=m_2$, $m_3=m_4$ or $m_1=m_4$, $m_2=m_3$, where $\{ m_1,m_2,m_3,m_4 \} =1,\dots ,M$, $\{ n_1,n_2,n_3,n_4 \} = 1,\dots ,N $, $d=( m_1-1 ) N+n_1$, $c=( m_2-1 ) N+n_2$, $a=( m_3-1 ) N+n_3$, $b=( m_4-1 ) N+n_4$, and $\overline{g}_{m_1k,n_1}$ is the $n_1$-th element of $\overline{\mathbf{g}}_{m_1k}$, respectively. $\mathrm{SINR}_{k}^{\mathrm{c},\mathrm{UatF},*}$ can also be computed in closed-form as 
$\overline{\mathrm{SINR}}_{k}^{\mathrm{c},\mathrm{UatF},*}=p_k\overline{\mathbf{r}}_k^H\overline{\mathbf{w}}_{k}^{*}$.
\end{thm}
\begin{IEEEproof}
The proof is provided in Appendix~\ref{app_centralized_obe_closed}.
\end{IEEEproof}

Based on the OBE matrices $\mathbf{W}_{k}^{*}$ designed as in this subsection, the CPU can implement the C-OBE combining scheme by plugging $\mathbf{W}_{k}^{*}$ into \eqref{Centrlized_Wg}.
\begin{rem}\label{C-OBE-WO-Phase}
Note that the considered random phase-shifts considered result in the phenomenon that some cross terms from different APs or different UEs become 0. When the phase-shifts are assumed to be perfectly mitigated through all processing phases in the system, so the channel model in Sec.~\ref{channel} reduces to $\mathbf{g}_{mk}=\overline{\mathbf{g}}_{mk}+\check{\mathbf{g}}_{mk}$, the closed-form results in Theorem~\ref{thm_centralized_closed} and Theorem~\ref{Centralized_OBE_closed} would have different forms. More specifically, we obtain the expressions in Theorem~\ref{thm_centralized_closed}, but using $\overline{\mathbf{G}}_{kk}=\overline{\mathbf{g}}_{k}^{\prime}\overline{\mathbf{g}}_{k}^{\prime,H}$ with $\overline{\mathbf{g}}_{k}^{\prime}\triangleq [\overline{\mathbf{g}}_{1k}^{T},\dots ,\overline{\mathbf{g}}_{Mk}^{T}]^T$,
\begin{equation}
\begin{aligned}\notag
\varepsilon _{kl}&=\sqrt{p_kp_l}\tau _p\mathrm{tr}( \mathbf{W}_{k}^{H}\overline{\mathbf{G}}_{lk} ) \mathrm{tr}( \mathbf{W}_k\check{\mathbf{R}}_k\mathbf{\Psi }_{k}^{-1}\check{\mathbf{R}}_l )\\
&+\sqrt{p_kp_l}\tau _p\mathrm{tr}( \mathbf{W}_k\overline{\mathbf{G}}_{kl} ) \mathrm{tr}( \mathbf{W}_{k}^{H}\check{\mathbf{R}}_l\mathbf{\Psi }_{k}^{-1}\check{\mathbf{R}}_k ) \\
&+p_kp_l\tau _{p}^{2}| \mathrm{tr}( \mathbf{W}_{k}^{H}\check{\mathbf{R}}_l\mathbf{\Psi }_{k}^{-1}\check{\mathbf{R}}_k ) |^2
\end{aligned}
\end{equation}
with  $\overline{\mathbf{G}}_{lk}=\overline{\mathbf{g}}_{l}^{\prime}\overline{\mathbf{g}}_{k}^{\prime,H}$, and $\omega _k=0$, respectively. Furthermore, the expressions in Theorem~\ref{Centralized_OBE_closed} hold but using $\mathbf{\Gamma }_k=\sum_{l=1}^K{p_l[ ( \overline{\mathbf{G}}_{kk}^{T}\otimes \overline{\mathbf{G}}_{ll} ) +( \overline{\mathbf{G}}_{kk}^{T}\otimes \check{\mathbf{R}}_l ) +( \widehat{\mathbf{R}}_{k}^{T}\otimes \overline{\mathbf{G}}_{ll} ) }+( \widehat{\mathbf{R}}_{k}^{T}\otimes \check{\mathbf{R}}_l ) ]+\sum_{l\in \mathcal{P} _k}{p_l[ \sqrt{p_kp_l}\tau _p\overline{\mathbf{g}}_{lk}\tilde{\mathbf{r}}_{lk}^{H}+\sqrt{p_kp_l}\tau _p\tilde{\mathbf{r}}_{lk}\overline{\mathbf{g}}_{lk}^{H}}+p_kp_l\tau _{p}^{2}\tilde{\mathbf{r}}_{lk}\tilde{\mathbf{r}}_{lk}^{H} ]-p_k\overline{\mathbf{r}}_k\overline{\mathbf{r}}_{k}^{H}+\sigma ^2( \overline{\mathbf{R}}_{k}^{T}\otimes \mathbf{I}_{MN} ) $, where $\overline{\mathbf{g}}_{lk}=\mathrm{vec}( \overline{\mathbf{G}}_{lk} ) $. These results can be easily derived following the same methods as in Appendix~\ref{app_cenralized_closed} and Appendix~\ref{app_centralized_obe_closed}. As observed, if the phase-shifts are perfectly mitigated, some cross terms from different APs or different UEs become non-zero, such as $\overline{\mathbf{G}}_{lk}$. Note that the results for the distributed processing scheme, including the achievable SE, DG-OBE, and DL-OBE expressions in scenarios with perfectly mitigated phase-shifts, can also be derived using the same approach as this remark. Therefore, they are omitted in the subsequent discussion for brevity.
\end{rem}

\vspace*{-0.4cm}

\subsection{DG-OBE Combining Scheme Design}\label{Sec-DG-OBE}
In this subsection, we study the DG-OBE combining scheme design to maximize \eqref{SE_LSFD}. We define $\mathbf{w}_{k}^{\mathrm{d}}=[ \mathbf{w}_{1k}^{T},\mathbf{w}_{2k}^{T},\dots ,\mathbf{w}_{Mk}^{T} ] ^T\in \mathbb{C} ^{MN^2}$, where $\mathbf{w}_{mk}=\mathrm{vec}( \mathbf{W}_{mk})\in \mathbb{C} ^{N^2}$. To separate the optimization of $\{ \mathbf{W}_{mk}:m=1,\dots ,M \} $ and $\mathbf{a}_k$, we first assume all elements in $\mathbf{a}_k$ to be equal and thus, $\mathrm{SINR}_{k}^{\mathrm{LSFD}}$ in \eqref{SINR_LSFD} reduces to \eqref{SINR_Distributed} on the top of the previous page. Based on \eqref{SINR_Distributed}, we can optimize the distributed BE matrices $\{ \mathbf{W}_{mk}:m=1,\dots ,M \} $ to maximize the achievable SE performance for UE $k$ in \eqref{SINR_Distributed} as follows.

\begin{coro}\label{DG-OBE-Monte}
The optimal matrices $\{ \mathbf{W}_{mk}^{*}:m=1,\dots ,M \} $, maximizing the effective SINR for the distributed processing scheme in \eqref{SINR_Distributed}, can be computed as $\mathbf{W}_{mk}^{*}=\mathrm{vec}^{-1}( \mathbf{w}_{mk}^{\mathrm{d},*})\in \mathbb{C} ^{N \times N}$ with $\mathbf{w}_{mk}^{\mathrm{d},*}=[ \mathbf{w}_{k}^{\mathrm{d},*} ] _{( m-1 ) N^2+1:mN^2}\in \mathbb{C} ^{N^2}$ and
\setcounter{equation}{26}
\begin{equation}
\begin{aligned}\label{DG_OBE_W_Monte}
&\mathbf{w}_{k}^{\mathrm{d},*}=\\
&\left( \sum_{l=1}^K{p_l\mathbb{E} \{ \mathbf{z}_{kl}\mathbf{z}_{kl}^{H} \} -p_k\mathbb{E} \{ \mathbf{h}_k \} \mathbb{E} \{ \mathbf{h}_k \} ^H +\sigma ^2\mathbf{\Theta }_{k}}\right) ^{-1}\mathbb{E} \{ \mathbf{h}_k \},
\end{aligned}
\end{equation}
where $\mathbf{z}_{kl}=[ \mathrm{vec}( \mathbf{g}_{1l}\widehat{\mathbf{g}}_{1k}^{H} ) ^T,\dots ,\mathrm{vec}( \mathbf{g}_{Ml}\widehat{\mathbf{g}}_{Mk}^{H} ) ^T ]^T\in \mathbb{C} ^{MN^2}$, $\mathbf{h}_k=[ \mathrm{vec}( \mathbf{g}_{1k}\widehat{\mathbf{g}}_{1k}^{H} ) ^T,\dots ,\mathrm{vec}( \mathbf{g}_{Mk}\widehat{\mathbf{g}}_{Mk}^{H} ) ^T ] ^T\in \mathbb{C} ^{MN^2}$, and $\mathbf{\Theta }_{k}=\mathrm{diag}\{ ( \mathbb{E} \{ \widehat{\mathbf{g}}_{1k}\widehat{\mathbf{g}}_{1k}^{H} \} ^T\otimes \mathbf{I}_N ) ,\dots ,( \mathbb{E} \{ \widehat{\mathbf{g}}_{Mk}\widehat{\mathbf{g}}_{Mk}^{H} \} ^T\otimes \mathbf{I}_N ) \} \in \mathbb{C} ^{MN^2\times MN^2}$. When $\{ \mathbf{W}_{mk}^{*}:m=1,\dots ,M \} $ are applied, the SINR in \eqref{SINR_Distributed} can achieve the maximum value as $\widetilde{\mathrm{SINR}}_{k}^{\mathrm{LSFD},*}=p_k\mathbb{E} \{\mathbf{h}_k\}^H\mathbf{w}_{k}^{\mathrm{d},*}$.
\end{coro}
\begin{IEEEproof}
The proof is provided in Appendix~\ref{app_distributed_obe}.
\end{IEEEproof}

We can also derive the closed-form results for $\mathbf{w}_{k}^{\mathrm{d},*}$ and $ \mathrm{SINR}_{k}^{\mathrm{d},*}$ as follows.

\begin{thm}\label{Distributed_OBE_closed}
The closed-form results of $\mathbf{w}_{k}^{\mathrm{d},*}$ can be computed as $\overline{\mathbf{w}}_{k}^{\mathrm{d},*}=\mathbf{\Lambda }_{k}^{-1}\overline{\mathbf{h}}_k$, where
$\mathbf{\Lambda }_k=\sum_{l=1}^K{\mathbf{\Lambda }_{kl}^{( 1 )} }+\sum_{l\in \mathcal{P} _k}{( \mathbf{\Lambda }_{kl}^{( 2 )}+\mathbf{\Lambda }_{kl}^{( 3 )} )}-p_k\overline{\mathbf{h}}_k\overline{\mathbf{h}}_{k}^{H}+\sigma ^2\mathbf{\Lambda }_{k}^{( 4 )} \in \mathbb{C} ^{MN^2 \times MN^2}$ and $\overline{\mathbf{h}}_k=[ \overline{\mathbf{h}}_{1k}^{T},\dots ,\overline{\mathbf{h}}_{Mk}^{T} ] ^T\in \mathbb{C} ^{MN^2}$ with $\overline{\mathbf{h}}_{mk}=\mathrm{vec}( \overline{\mathbf{R}}_{mk} ) \in \mathbb{C} ^{N^2}$. Besides, $\mathbf{\Lambda }_{kl}^{( 1 )}=\mathrm{diag}( \mathbf{\Lambda }_{1kl}^{( 1 )},\dots ,\mathbf{\Lambda }_{Mkl}^{( 1 )} ) $, $\mathbf{\Lambda }_{kl}^{( 2 )}=\mathrm{diag}( \mathbf{\Lambda }_{1kl}^{( 2 )},\dots ,\mathbf{\Lambda }_{Mkl}^{( 2 )} ) $, $\mathbf{\Lambda }_{k}^{( 4)}=\mathrm{diag}( \mathbf{\Lambda }_{1k}^{( 4 )},\dots ,\mathbf{\Lambda }_{Mk}^{( 4 )} )$ are all diagonal block matrices with dimension $\mathbb{C} ^{MN^2\times MN^2}$, where $\mathbf{\Lambda }_{mkl}^{( 1 )}=p_l[ ( \overline{\mathbf{G}}_{mkk}^{T}\otimes \overline{\mathbf{G}}_{mll} ) +( \overline{\mathbf{G}}_{mkk}^{T}\otimes \check{\mathbf{R}}_{ml} ) +( \widehat{\mathbf{R}}_{mk}^{T}\otimes \overline{\mathbf{G}}_{mll} ) +( \widehat{\mathbf{R}}_{mk}^{T}\otimes \check{\mathbf{R}}_{ml} ) ] $, 
\begin{equation}
\begin{aligned}
&\mathbf{\Lambda }_{mkl}^{\left( 2 \right)}=p_kp_{l}^{2}\tau _{p}^{2}\tilde{\mathbf{r}}_{mlk}\tilde{\mathbf{r}}_{mlk}^{H}+\\
&\begin{cases}
	\mathbf{0},\,\,\mathrm{if}\,\,l\in \mathcal{P} _k\backslash \left\{ k \right\} \,\,\\
	p_{k}^{2}\tau _p\overline{\mathbf{g}}_{mkk}\tilde{\mathbf{r}}_{mkk}^{H}+p_{k}^{2}\tau _p\tilde{\mathbf{r}}_{mkk}\overline{\mathbf{g}}_{mkk}^{H}, \mathrm{if}\,\,l=k\\
\end{cases}
\end{aligned}
\end{equation}
with $\tilde{\mathbf{r}}_{mlk}=\mathrm{vec}( \check{\mathbf{R}}_{ml}\mathbf{\Psi }_{mk}^{-1}\check{\mathbf{R}}_{mk} )$, $\overline{\mathbf{g}}_{mkk}=\mathrm{vec}( \overline{\mathbf{G}}_{mkk} ) $, and $\mathbf{\Lambda }_{mk}^{( 4 )}= \overline{\mathbf{R}}_{mk}^{T}\otimes \mathbf{I}_N $. Moreover, we have $\mathbf{\Lambda }_{kl}^{\left( 3 \right)}=p_l( \check{\mathbf{B}}_{lk}-\overline{\mathbf{B}}_{lk}) $, where $\check{\mathbf{B}}_{lk}=\mathbf{b}_{lk}\mathbf{b}_{lk}^{H}$ and $\overline{\mathbf{B}}_{lk}=\mathrm{diag}( \overline{\mathbf{B}}_{1lk},\dots ,\overline{\mathbf{B}}_{Mlk} ) \in \mathbb{C} ^{MN^2\times MN^2}$ with 
$\mathbf{b}_{lk}=\left[ \mathbf{b}_{1lk}^{T},\dots ,\mathbf{b}_{Mlk}^{T} \right] ^T$, $\overline{\mathbf{B}}_{mlk}=\mathbf{b}_{mlk}\mathbf{b}_{mlk}^{H}$, and $\mathbf{b}_{mlk}=\mathrm{vec}( \mathbf{B}_{mlk} )$. The maximum value of SINR can also be computed in closed-form as $\overline{\widetilde{\mathrm{SINR}}}_{k}^{\mathrm{LSFD},*}=p_k\overline{\mathbf{h}}_{k}^H\overline{\mathbf{w}}_{k}^{\mathrm{d},*}$.
\end{thm}
\begin{IEEEproof}
We provide the proof in Appendix~\ref{app_distributed_obe_closed}.
\end{IEEEproof}

\begin{rem}\label{Distributed_Implement}
As observed in Corollary~\ref{DG-OBE-Monte} and Theorem~\ref{Distributed_OBE_closed}, the optimal design of $\mathbf{W}_{mk}$ relies on global channel statistics information. However, given that channel statistics remain constant over a long period of time, it is feasible to implement the optimal design of $\mathbf{W}_{mk}$ locally at each AP. To facilitate this, the necessary channel statistics can be transmitted from other APs to each AP via fronthaul links for each realization of the AP/UE locations. Each AP can implement the DG-OBE combining scheme by plugging $\mathbf{W}_{mk}^{*}$ into \eqref{local_generalized}. It is worth noting that the DG-OBE combining scheme allows for the distributed implementation at each AP by utilizing $\mathbf{W}_{mk}^{*}$ based on global channel statistics and local instantaneous channel estimates $\widehat{\mathbf{g}}_{mk}$ at each AP. The results in Theorem~\ref{Distributed_Implement} can reduce to so-called ``GMR" combining as \cite[Eq. (15)]{polegre2021pilot} when $\overline{\mathbf{G}}_{mkl}=\mathbf{0}$ and $N=1$.
\end{rem}
\vspace*{-0.2cm}

Note that the optimal design of $\mathbf{W}_{mk}$ is implemented based on the performance analysis framework in \eqref{SINR_Distributed}. It is also available for the LSFD processing scheme discussed in Sec.~\ref{distributed} by plugging the derived DG-OBE schemes into results in Sec.~\ref{distributed}. We notice that, when applying the BE combining scheme in \eqref{local_generalized} at each AP, the final decoding data in the LSFD processing scheme can be denoted by
$\check{x}_k=( \sum_{m=1}^M{a_{mk}^{*}\widehat{\mathbf{g}}_{mk}^{H}\mathbf{W}_{mk}^{H}\mathbf{g}_{mk}} ) x_k+\sum_{l\ne k}^K{\sum_{m=1}^M{a_{mk}^{*}\widehat{\mathbf{g}}_{mk}^{H}\mathbf{W}_{mk}^{H}\mathbf{g}_{ml}x_l}}+\mathbf{n}_{k}^{\prime}$. The LSFD coefficient $a_{mk}$ can be regarded as an additional scaling factor for $\mathbf{W}_{mk}$.

\begin{rem}\label{noneed_LSFD}
For the conventional LSFD processing scheme with the arbitrary $\mathbf{W}_{mk}$, the LSFD coefficients can be optimized as in \eqref{optimal_lsfd_eq}. 
However, when DG-OBE scheme in Corollary~\ref{DG-OBE-Monte} is applied at each AP, the final decoding step at the CPU using the optimal LSFD coefficients becomes unnecessary since it can be regarded as that BE matrices with additional LSFD coefficients $\tilde{\mathbf{W}}_{mk}=a_{mk}\mathbf{W}_{mk}$ have been unifiedly optimized to maximize the achievable SE in \eqref{SE_LSFD}. 
So there will be no further SE performance improvement from applying the optimal LSFD strategy if the DG-OBE combining scheme in Corollary~\ref{DG-OBE-Monte} is applied. This observation will also be substantiated in the simulation results, like Fig.~\ref{6}.
In other words, the optimal LSFD scheme design in \eqref{optimal_lsfd_eq} can be considered to be a special case of DG-OBE scheme, where $\mathbf{W}_{mk}$ is fixed except for a scaling factor. Moreover, the DG-OBE scheme in Corollary~\ref{DG-OBE-Monte} reduces to the optimal LSFD scheme in \eqref{optimal_lsfd_eq} with the L-MR combining under the case with single-antenna APs.
\end{rem}

\begin{rem}
Note that the performance analysis framework for the distributed processing scheme in \eqref{SE_LSFD} holds for arbitrary local combining vectors $\mathbf{v}_{mk}$. The distributed OBE scheme $\mathbf{v}_{mk}=\mathbf{W}_{mk}\widehat{\mathbf{g}}_{mk}$ introduced in this part can be equivalently viewed as the combination of the L-MR combining scheme $\overline{\mathbf{v}}_{mk}=\widehat{\mathbf{g}}_{mk}$ and the weighting matrix $\mathbf{W}_{mk}$. Furthermore, a more generalized decoding scheme can be extended from the distributed OBE scheme as $\mathbf{v}_{mk}=\mathbf{W}_{mk}\overline{\mathbf{v}}_{mk}$, where $\overline{\mathbf{v}}_{mk}$ denotes an arbitrary local combining vector. Following a similar approach as in Corollary~\ref{DG-OBE-Monte}, the weighting matrix $\mathbf{W}_{mk}$ can be optimized to maximize the achievable SE performance by replacing $\widehat{\mathbf{g}}_{mk}$ in \eqref{DG_OBE_W_Monte} by $\overline{\mathbf{v}}_{mk}$, which can be called as the OBE weighting strategy. Different from Corollary~\ref{DG-OBE-Monte}, the optimal OBE weighting strategy holds for an arbitrary local combining vector $\overline{\mathbf{v}}_{mk}$. Moreover, different from the local channel estimates-based combining scheme, the channel state information from other APs can also be involved in the design of $\overline{\mathbf{v}}_{mk}$. One possible scheme is the team MMSE combining scheme, where each AP can involve the channel estimates from other APs in a sequential manner to design the local combining scheme as in \cite{miretti2022team}. By constructing these local combining schemes with the aid of the channel state information from other APs or neighboring APs, we can derive the corresponding OBE weighting strategy. This combination is anticipated to further enhance the SE performance compared to the scheme with the OBE weighting scheme and local combining scheme with only local channel state information at each AP.
\end{rem}

\vspace{-0.5cm}

\subsection{DL-OBE Combining Scheme Design}\label{Sec-DL-OBE}
\vspace*{-0.2cm}
Note that the DG-OBE combining matrices introduced in Sec.~\ref{Sec-DG-OBE} is designed based on the global channel statistics information. To realize the distributed design at each AP, the required statistics information needs to be delivered from the other APs via fronthaul links. In practice, the user scheduling could change rapidly so then one anyway needs to transfer statistics often. To relieve the often statistics transfer when designing the OBE scheme, in this subsection, we study the DL-OBE combining scheme design. As evident from its name, only the local channel statistics information at each AP is needed to locally design the OBE matrices in this combining scheme.

At each AP $m$, when only its local information is applied to decode the signal and compute the achievable performance for one particular UE $k$, the achievable SE for UE $k$ at AP $m$ based on the UatF capacity bound can be denoted as $\mathrm{SE}_{mk}^{\mathrm{d}}=\frac{\tau _c-\tau _p}{\tau _c}\log _2(1+\mathrm{SINR}_{mk}^{\mathrm{d}})$, where 
\begin{equation}
\begin{aligned}\label{SINR_Local}
&\mathrm{SINR}_{mk}^{\mathrm{d}}=\\
&\frac{p_k | \mathbb{E} \{ \mathbf{v}_{mk}^{H}\mathbf{g}_{mk} \} |^2}{\sum\limits_{l=1}^K{p_l\mathbb{E} \{ | \mathbf{v}_{mk}^{H}\mathbf{g}_{ml} |^2 \}}-p_k| \mathbb{E} \{ \mathbf{v}_{mk}^{H}\mathbf{g}_{mk} \} |^2+\sigma ^2\mathbb{E} \{ \| \mathbf{v}_{mk} \| ^2 \}},
\end{aligned}
\end{equation}
which can be easily proven based on the method of the centralized processing scheme. Based on \eqref{SINR_Local}, we can derive the optimal  $\mathbf{W}_{mk}$, that maximizes \eqref{SINR_Local} as follows.

\begin{coro}\label{D_OBE_UatF_Local}
The optimal $\mathbf{W}_{mk}$, maximizing the effective SINR for UE $k$ at AP $m$ as \eqref{SINR_Local}, can be denoted by $\mathbf{W}_{mk}^{*}=\mathrm{vec}^{-1}( \mathbf{w}_{mk}^{*} ) $ with the maximum value $\mathrm{SINR}_{mk}^{\mathrm{d},*}=p_k\mathrm{vec}( \mathbb{E} \{ \mathbf{g}_{mk}\widehat{\mathbf{g}}_{mk}^{H} \} ) ^H\mathbf{w}_{mk}^{*}$, where 
$\mathbf{w}_{mk}^{*}=[ \sum_{l=1}^K{p_l\mathbb{E} \{  ( \widehat{\mathbf{g}}_{mk}\widehat{\mathbf{g}}_{mk}^{H} ) ^T\otimes ( \mathbf{g}_{ml}\mathbf{g}_{ml}^{H} )  \}}-p_k\mathrm{vec}( \mathbb{E} \{ \mathbf{g}_{mk}\widehat{\mathbf{g}}_{mk}^{H} \} ) \mathrm{vec}( \mathbb{E} \{ \mathbf{g}_{mk}\widehat{\mathbf{g}}_{mk}^{H} \} ) ^H+( \mathbb{E} \{ \widehat{\mathbf{g}}_{mk}\widehat{\mathbf{g}}_{mk}^{H} \} ^T\otimes \mathbf{I}_{N} ) ] ^{-1}\mathrm{vec}( \mathbb{E} \{ \mathbf{g}_{mk}\widehat{\mathbf{g}}_{mk}^{H} \} )$.
\end{coro}
\begin{IEEEproof}
The proof can be easily derived based on the similar methods as in Corollary~\ref{C_OBE_UatF}.
\end{IEEEproof}

The results in Corollary~\ref{D_OBE_UatF_Local} can also be computed in closed-form as follows.

\begin{thm}\label{Dis_OBE_local_closed}
We can compute $\mathbf{w}_{k}^{*}$ in closed-form as
\setcounter{equation}{29}
\begin{equation}\label{Dis_obe_local_vector}
\overline{\mathbf{w}}_{mk}^{*}=\mathbf{\Gamma }_{mk}^{-1}\overline{\mathbf{r}}_{mk},
\end{equation}
where $\mathbf{\Gamma }_{mk}=\sum\nolimits_{l=1}^K{p_l[ ( \overline{\mathbf{G}}_{mkk}^{T}\otimes \overline{\mathbf{G}}_{mll} ) +( \overline{\mathbf{G}}_{mkk}^{T}\otimes \check{\mathbf{R}}_{ml} ) }+( \widehat{\mathbf{R}}_{mk}^{T}\otimes \overline{\mathbf{G}}_{mll} ) +( \widehat{\mathbf{R}}_{mk}^{T}\otimes \check{\mathbf{R}}_{ml} ) ]+\sum\nolimits_{l\in \mathcal{P} _k}{p_kp_{l}^2\tau _{p}^{2}\tilde{\mathbf{r}}_{mlk}\tilde{\mathbf{r}}_{mlk}^{H} }+p_{k}^{2}\tau _p\overline{\mathbf{g}}_{mkk}\tilde{\mathbf{r}}_{mkk}^{H}+p_{k}^{2}\tau _p\tilde{\mathbf{r}}_{mkk}\overline{\mathbf{g}}_{mkk}^{H}-p_k\overline{\mathbf{r}}_{mk}\overline{\mathbf{r}}_{mk}^{H}+\sigma ^2( \overline{\mathbf{R}}_{mk}^{T}\otimes \mathbf{I}_N )\in \mathbb{C} ^{N^2 \times N^2}$. The closed-form result of $\mathrm{SINR}_{mk}^{\mathrm{d},*}$ is $\overline{\mathrm{SINR}}_{mk}^{\mathrm{d},\mathrm{UatF},*}=p_k\overline{\mathbf{r}}_{mk}^H\overline{\mathbf{w}}_{mk}^{*}$.
\end{thm}
\begin{IEEEproof}
This result is easily proved following the same approach as Theorem~\ref{Centralized_OBE_closed}
\end{IEEEproof}

\begin{table}[t!]
  \centering
  \fontsize{8}{9}\selectfont
  \caption{Computational complexity of considered schemes for each realization of the AP/UE locations.}
  \vspace*{-0.3cm}
  \label{Comparisons}
   \begin{tabular}{ !{\vrule width0.7 pt}  m{1.2 cm}<{\centering} !{\vrule width0.7pt}  m{3.1 cm}<{\centering} !{\vrule width0.7pt}  m{3.1cm}<{\centering} !{\vrule width0.7pt}}

    \Xhline{0.7pt}
         \bf Scheme & \bf Combining design  & \bf Precomputation based on statistics \cr
    \Xhline{0.7pt}
    C-MMSE & $\mathcal{O} ( M^3KN^3I_r+M^2KN^2I_r ) $ & $\mathcal{O} \left( MKN^3 \right)  $\cr\hline
    C-OBE & $\mathcal{O} ( M^2KN^2I_r ) $   & $\mathcal{O} ( M^6KN^6+M^4KN^4\lfloor K/\tau _p \rfloor )  $\cr\hline
    L-MMSE & $\mathcal{O} ( MKN^2I_r + MN^3I_r) $   & $\mathcal{O} \left( MKN^3 \right)$\cr\hline
    DG-OBE & $\mathcal{O} (MKN^2I_r) $ &$\mathcal{O} ( M^3KN^6+M^2KN^4\lfloor K/\tau _p \rfloor)  $\cr\hline
    DL-OBE &$\mathcal{O} (MKN^2I_r) $ & $\mathcal{O} (MKN^6+MKN^4\lfloor K/\tau _p \rfloor  )  $  \cr\hline
    LSFD & $-$ & $\mathcal{O} (MK^2N^3+M^3K) $  \cr\hline

    \Xhline{0.7pt}
    \end{tabular}
  \vspace*{-0.6cm}
\end{table}

Note that the performance analysis framework discussed in this part such as \eqref{SINR_Local} is motivated from those of small-cell networks \cite{[162]}, but relies on the UatF capacity bound. In the small-cell networks in \cite{[162]}, the standard capacity bound was applied to evaluate the performance as $\mathrm{SE}_{mk}^{\mathrm{d},\mathrm{stan}}=\frac{\tau _c-\tau _p}{\tau _c}\mathbb{E} \{ \log _2( 1+\mathrm{SINR}_{mk}^{\mathrm{d},\mathrm{stan}} ) \}$,
where
\begin{equation}
\begin{aligned}\label{SINR_Distributed_Standard}
&\mathrm{SINR}_{mk}^{\mathrm{d},\mathrm{stan}}=\\
&\frac{p_k| \mathbf{v}_{mk}^{H}\widehat{\mathbf{g}}_{mk} |^2}{\sum\limits_{l\ne k}^K{p_l| \mathbf{v}_{mk}^{H}\widehat{\mathbf{g}}_{ml} |^2}+\mathbf{v}_{mk}^{H}( \sum\limits_{l=1}^K{p_l\mathbf{C}_{ml}}+\sigma ^2\mathbf{I}_{N} ) ^{-1}\mathbf{v}_{mk}},
\end{aligned}
\end{equation}
is the instantaneous effective SINR for UE $k$ at AP $m$. When this standard bound is applied, the L-MMSE combining in \eqref{local_MMSE} can maximize \eqref{SINR_Distributed_Standard}. However, the L-MMSE combining is designed based on the instantaneous channel estimates. When the UatF capacity bound in \eqref{SINR_Local} is applied, the DL-OBE combining scheme introduced in this part can maximize \eqref{SINR_Local}.

\vspace*{-0.2cm}

\begin{rem}\label{Distributed_local}
When the DL-OBE combining scheme is applied at each AP, the distributed processing scheme in Sec.~\ref{distributed} with the LSFD processing scheme, instead of the fully local decoding scheme in \eqref{SINR_Local}, can be implemented to further enhance the SE performance for the networks. Thus, the DL-OBE combining scheme can be regarded as a heuristic one since, from the perspective of the decoding scheme in the whole network. And the fully local decoding scheme in \eqref{SINR_Local} is not applied for the decoding but only for the DL-OBE combining scheme design.
\end{rem}

In Table~\ref{Comparisons}, we summarize the computational complexity of the considered schemes in this paper, where $I_r$ denotes the number of channel realizations for each realizations of the AP/UE locations and $\lfloor K/\tau _p \rfloor$ denotes the average number of UEs sharing the same pilot signal. As observed, the OBE-based combining scheme design embraces much lower computational complexity than that of MMSE-based combining scheme design, albeit at the cost of higher precomputation complexity. This higher precomputation complexity is deemed acceptable since it relies solely on channel statistics and needs to be performed only once for each realization of the AP/UE locations.

However, in large-scale or highly dynamic networks, this precomputation complexity will become challenging. More specifically, for the large-scale or dense networks with large values of $\{M, K, N\}$, the precomputation complexity becomes quite large even if the computation occurs only once per major change in the AP/UE configuration. For highly dynamic networks with high user mobility or rapid changes in the AP/UE configurations, the channel statistics vary frequently, necessitating more frequent recalculations of the required precomputed matrices than those of the low mobility scenarios. These two important issues pose noticeable challenges for the practical implementation of the OBE schemes, especially the C-OBE scheme with the highest precomputation complexity. To facilitate the practical implementation of the proposed OBE schemes, here are three promising enablers:

\emph{(1) The distributed OBE schemes are advocated in large-scale or highly dynamic networks:} As will be comprehensively discussed in the simulation part, such as Fig.~\ref{All_Schemes}, the distributed OBE schemes, especially the DL-OBE scheme, show competitive advantage in the highly dynamic scenario with much smaller precomputation complexity than the C-OBE scheme. Thus, to balance the achievable performance and precomputation complexity, the distributed OBE schemes, especially the DL-OBE scheme, are highly advocated in the large-scale or highly dynamic networks.

\emph{(2) Low-complexity algorithms can be applied to enhance the feasibility of OBE schemes in practice:} Another promising solution is to apply some low-complexity algorithms to significantly reduce the precomputation complexity, such as the low-complexity matrix inversion algorithms. Taking the C-OBE scheme as an example, in \eqref{Optimal_centralized_obe}, for each UE, the major computational complexity of the C-OBE scheme comes from the $M^2N^2\times M^2N^2$ matrix inversion, which is $\mathcal{O} (M^6N^6)$. By applying low-complexity matrix inversion algorithms, such as the symmetric successive over relaxation (SSOR) algorithm \cite{xie2016low}, which can provide an approximate calculation of the matrix inversion, the computational complexity of this matrix inversion can be efficiently degraded to  $\mathcal{O} (M^4N^4N_{Iter})$ with $N_{Iter}$ being the number of iterations of the SSOR algorithm, which is usually set to a small value. Besides, other low-complexity algorithms are also advocated to further degrade the precomputation complexity, which is regarded as an important future direction.

\emph{(3) Machine learning can be utilized to empower the OBE design in highly dynamic networks:} To facilitate the OBE design in highly dynamic networks, where the statistics frequently vary due to the mobility and AP/UE configuration, the machine learning methodology can be utilized to help learn how the statistical parameters change in practice. Besides, another promising methodology is to learn how to estimate the statistics from received data signals, where the channel map can be applied to identify statistical parameters with data collected from previous users located on the same location \cite{chen2025channel}. Thus, the intelligent OBE design can be efficiently empowered by the machine learning methodology based on the learned statistical parameters.  Moreover, by learning the vectorization and vectorization reversion approaches for the optimization of BE matrices, the machine learning is also anticipated to directly construct the OBE matrices based on the statistics but avoid the computation of OBE vectors with high computational complexity.

\vspace{-0.2cm}

\begin{rem}
All the results are derived for Rician fading channels, but they also hold in the special case of Rayleigh fading channels. More specifically, by letting $\overline{\mathbf{g}}_{mk}=\mathbf{0}$, all results reduce to those with Rayleigh fading. Notably, Fig.~\ref{Rician_Region} verifies the results for the Rayleigh fading channel.
\end{rem}
\vspace{-0.2cm}

\begin{rem}\label{OBE-Pilot}
Compared to MMSE combining schemes, which rely on matrix inversion based on instantaneous channel estimates, OBE schemes are less sensitive to the quality of channel estimates due to their reliance on channel statistics-based OBE matrices. Therefore, OBE schemes are expected to remain robust in scenarios characterized by poor-quality channel estimates or severe pilot contamination. This observation will be further substantiated through numerical results, like Fig.~\ref{3}, Fig.~\ref{4}, Fig.~\ref{8}, and Fig.~\ref{Channel_Estimator}.
\end{rem}
\vspace{-0.5cm}

\section{Numerical Results}\label{num}
In this section, we consider a CF mMIMO network, where all APs and UEs are randomly distributed in a $1\times 1 \, \mathrm{km}^2$ area with the wrap-around scheme \cite{8187178}. Unless otherwise stated, we assume that all AP-UE pairs have LoS paths and the large-scale fading pathloss coefficient between AP $m$ and UE $k$ is denoted by $\beta_{mk}$, which is modeled based on the COST 321 Walfish-Ikegami model in \cite[Eq. (24)]{wang2020uplink}.
The large-scale fading coefficient for the LoS and NLoS components between AP $m$ and UE $k$ can be denoted as
$\beta _{mk}^{\mathrm{LoS}}=\frac{\kappa _{mk}}{\kappa _{mk}+1}\beta _{mk}$ and $\beta _{mk}^{\mathrm{NLoS}}=\frac{1}{\kappa _{mk}+1}\beta _{mk}$, respectively, where $\kappa _{mk}=10^{1.3-0.003d_{mk}}$ is the Rician $\kappa$ factor between AP $m$ and UE $k$ with $d_{mk}$ is the distance between AP $m$ and UE $k$ (taking the $11 \, \mathrm{m}$ height difference into account). We consider that each AP is equipped with a uniform linear array (ULA) with $N$ antennas. The $n$-th element for $\overline{\mathbf{g}}_{mk}$ is 
$[\overline{\mathbf{g}}_{mk}]_n=\sqrt{\beta _{mk}^{\mathrm{LoS}}}e^{j2\pi (n-1)\sin \theta _{mk}{{\lambda}/{2}}}$, where $\lambda$ is the wavelength and $\theta _{mk}$ is the angle of arrival between UE $k$ and AP $m$. Note that the spatial correlation matrix 
$\check{\mathbf{R}}_{mk} $ is modeled based on the Gaussian local scattering model in \cite[Sec. 2.6]{8187178}. Moreover, we have $\tau _c=200
$, $\sigma ^2=-94 \, \mathrm{dBm}$, $p_k=200 \, \mathrm{mW}$, and the bandwidth is $20 \, \mathrm{MHz}$. The pilot allocation strategy follows from that of \cite{wang2020uplink}. The sources of randomness embrace the random AP/UE locations and random fading realizations. More specifically, dozens or even hundreds of realizations of AP/UE locations are simulated and for each realization of the AP/UE locations, $1000$ channel realizations are considered. For the distributed processing scheme, the UatF capacity bound-based analysis framework is applied to evaluate the achievable SE performance. For the centralized processing scheme, here are two promising capacity bounds to evaluate the achievable SE performance: UatF capacity bound based on the channel statistics in \eqref{SE_centrlized_UatF} and the standard capacity bound based on the instantaneous channel state information in \cite[Proposition 1]{[162]}. We have mentioned which bound is applied in the following statements.

In Fig.~\ref{1}, we investigate the average SE per UE against the number of antennas per AP ($N$) for the fully centralized processing (FCP) scheme over different combining schemes. To validate the matching accuracy of the closed-form results for the FCP scheme with the Monte-Carlo simulation, we apply the UatF bound in \eqref{SE_centrlized_UatF} to evaluate the SE performance in this figure. As observed, with only channel statistics applied to design the C-OBE matrices, this scheme can achieve much better SE than the C-MR combining scheme. When $N=4$, the C-OBE combining can achieve $113.5 \%$ SE improvement over the C-MR combining. More importantly, for $N=4$, there is only $33 \%$ SE performance gap between the C-OBE and global optimal C-MMSE combining schemes. For the C-OBE and C-MR combining, markers ``$\circ$" generated by the closed-form results in Th.~\ref{thm_centralized_closed} and Th.~\ref{Centralized_OBE_closed} overlap with the curves generated by the Monte-Carlo simulation.

\begin{figure}[t]
\centering
\includegraphics[scale=0.45]{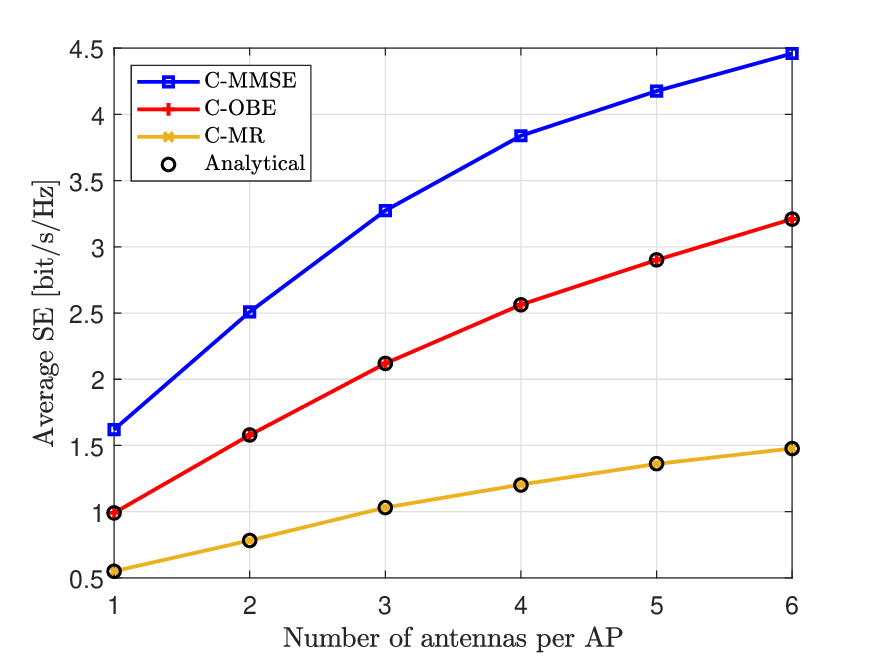}
\vspace{-0.37cm}
\caption{Average SE per UE measured by the UatF bound against the number of antennas per AP for the FCP scheme over different combining schemes with $M=20$, $K=20$, and $\tau_p=1$. \label{1}}
\vspace{-0.5cm}
\end{figure}

\begin{figure}[t]
\centering
\includegraphics[scale=0.45]{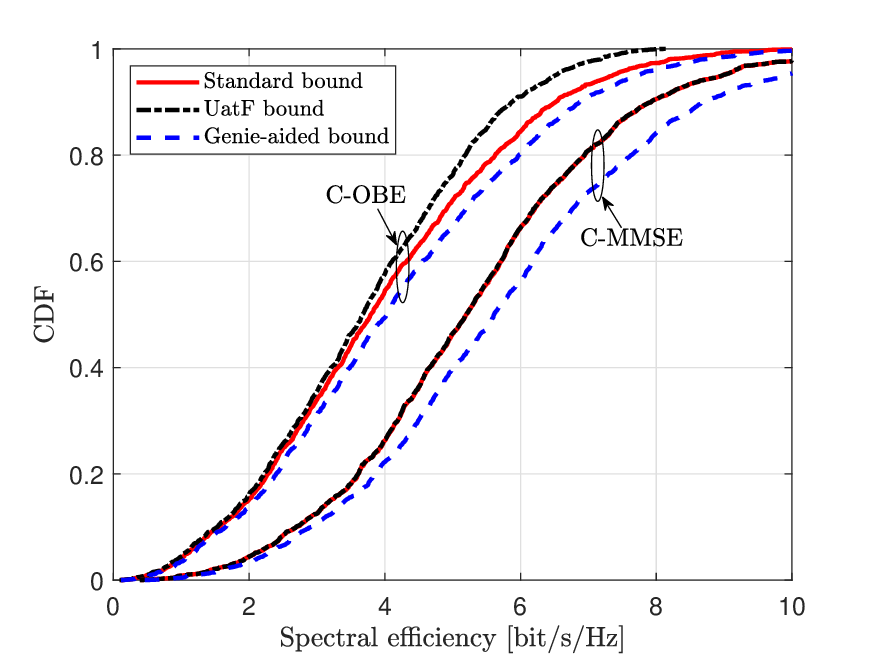}
\vspace{-0.37cm}
\caption{CDF of the SE per UE measured by the standard, UatF, and genie-aided capacity bounds for the FCP scheme with $M=20$, $K=10$, $N=4$, and $\tau_p=1$.\label{2}}
\vspace{-0.5cm}
\end{figure}

\begin{figure}[t]
\centering
\includegraphics[scale=0.45]{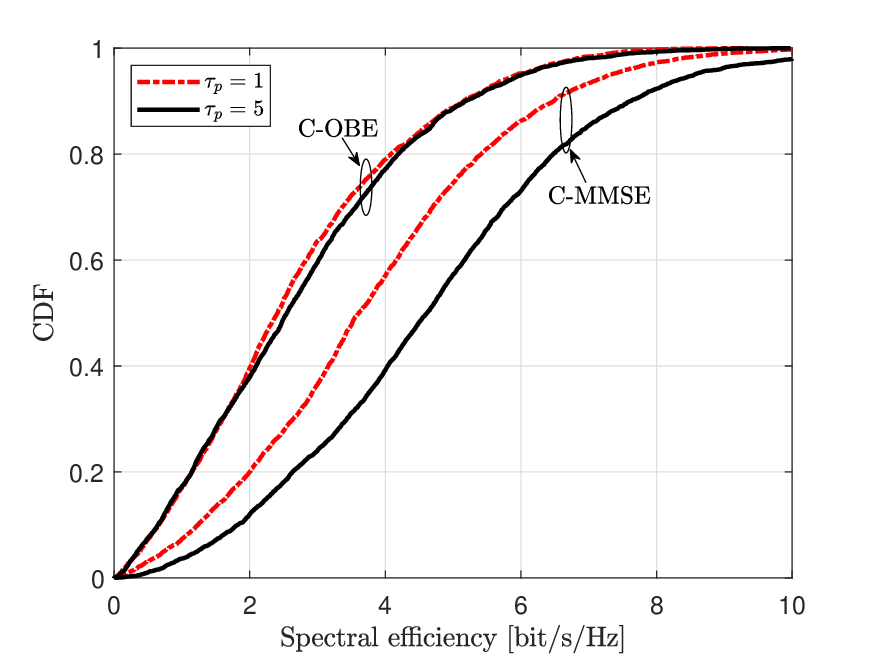}
\vspace{-0.37cm}
\caption{CDF of the SE per UE measured by the standard bound for the FCP scheme with different numbers of pilot signals $\tau_p$ over $M=20$, $K=20$ and $N=4$.\label{3}}
\vspace{-0.5cm}
\end{figure}

\begin{figure}[t]
\centering
\includegraphics[scale=0.45]{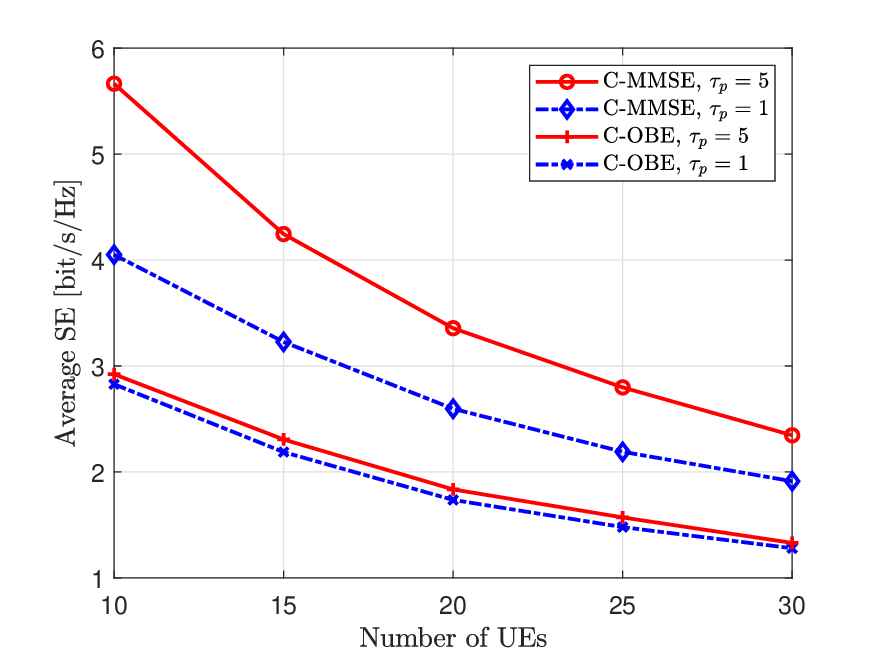}
\vspace{-0.37cm}
\caption{Average SE per UE measured by the standard bound against the number of UEs for the FCP scheme with $M=20$, $N=2$, and $\tau_p=\{1,5\}$. \label{4}}
\vspace{-0.5cm}
\end{figure}

Fig.~\ref{2} shows the cumulative distribution function (CDF) curves of the SE per UE, measured by the UatF capacity bound in \eqref{SE_centrlized_UatF}, the standard capacity bound in \cite[Proposition 1]{[162]}, and the genie-aided bound\footnote{Note that the so-called genie-aided bound is the one that applies the perfect channel knowledge to compute the achievable SE, which is an upper value for the SE.} in \cite[Corollary 5.9]{demir2021foundations} for the FCP scheme. We observe that when the C-MMSE combining scheme is applied, the performance gap between the SE performance evaluated by the UatF bound and that evaluated by the standard bound is negligible. This finding demonstrates that the UatF bound, which we utilized to design the C-OBE combining, is also efficient to depict the achievable SE performance. Moreover, there would be a further SE improvement by applying the standard capacity bound to evaluate the SE performance for the networks with the C-OBE combining scheme applied, compared with that of the UatF bound applied scenario. For instance, about $6.9 \%$ average SE improvement can be achieved. Thus, in subsequent discussions, unless otherwise stated, to fully explore the potentials of the C-OBE scheme, we will utilize the standard capacity bound as \cite[Proposition 1]{[162]} to evaluate the SE performance for the FCP scheme. Meanwhile, there are about $8.1 \%$ and $5.3 \%$ average SE performance gaps between the SE performance evaluated by the standard bound and the genie-aided bound for the C-MMSE and C-OBE combining schemes, respectively. These gaps are acceptable since the genie-aided bound neglects the effect of both the bounding error and channel estimation error.

Fig.~\ref{3} shows the CDF curves of the SE per UE for the FCP scheme with different numbers of pilot signals. $\tau_p=1$ denotes that all $K$ UEs are assigned with the same pilot signal, where the pilot contamination is most severe. It can be found that the SE performance gap between the C-OBE and C-MMSE combining schemes for $\tau_p=1$ is much smaller than that of $\tau_p=5$. For instance, only $29 \%$ average SE performance gap between the C-OBE and C-MMSE combining schemes for $\tau_p=1$, which is much smaller than $41.3 \%$ for $\tau_p=5$. This observation demonstrates that the C-OBE scheme maintains robustness with severe pilot contamination, which is also clarified in Remark~\ref{OBE-Pilot}. However, when the pilot contamination is moderate, the C-MMSE combining can achieve significantly better performance than the C-OBE combining with the aid of high-quality channel estimates.

\begin{figure}[t]
\centering
\includegraphics[scale=0.45]{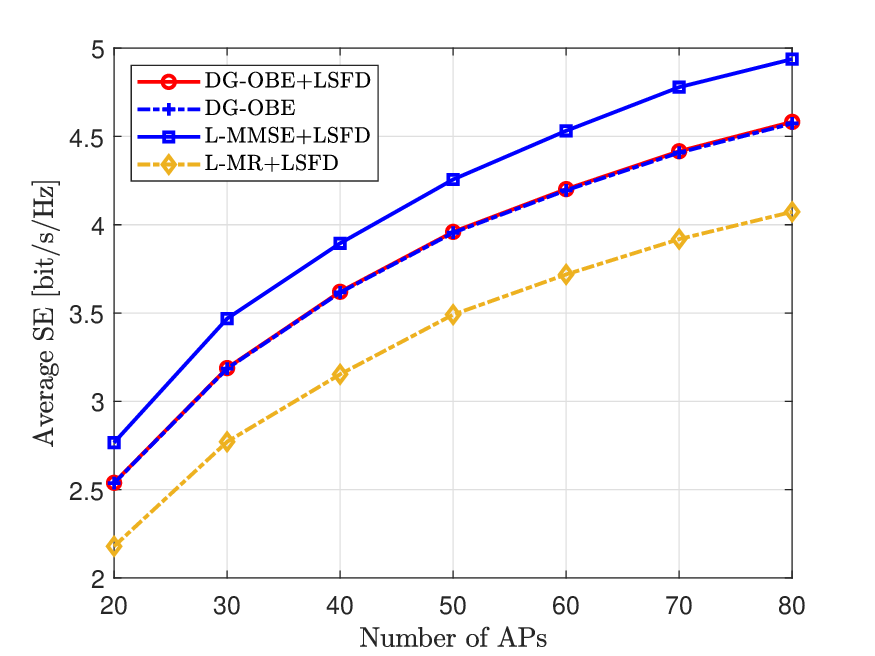}
\vspace{-0.37cm}
\caption{Average SE per UE versus the number of APs over the EWDP and LSFD schemes with $K=20$, $N=4$, and $\tau_p=1$.\label{6}}
\vspace{-0.5cm}
\end{figure}

\begin{figure}[t]
\centering
\includegraphics[scale=0.45]{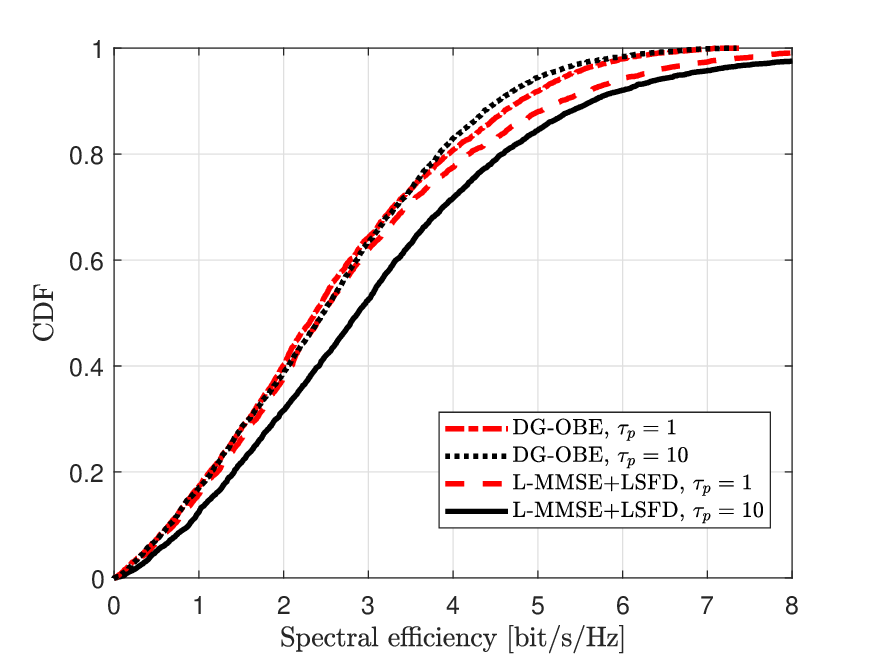}
\vspace{-0.37cm}
\caption{CDF of the SE per UE for L-MMSE+LSFD and DG-OBE schemes with different numbers of pilot signals $\tau_p$ over $M=20$, $K=20$, $N=4$, and $\tau_p=\{1,10\}$.\label{8}}
\vspace{-0.5cm}
\end{figure}

\begin{figure}[t]
\centering
\includegraphics[scale=0.45]{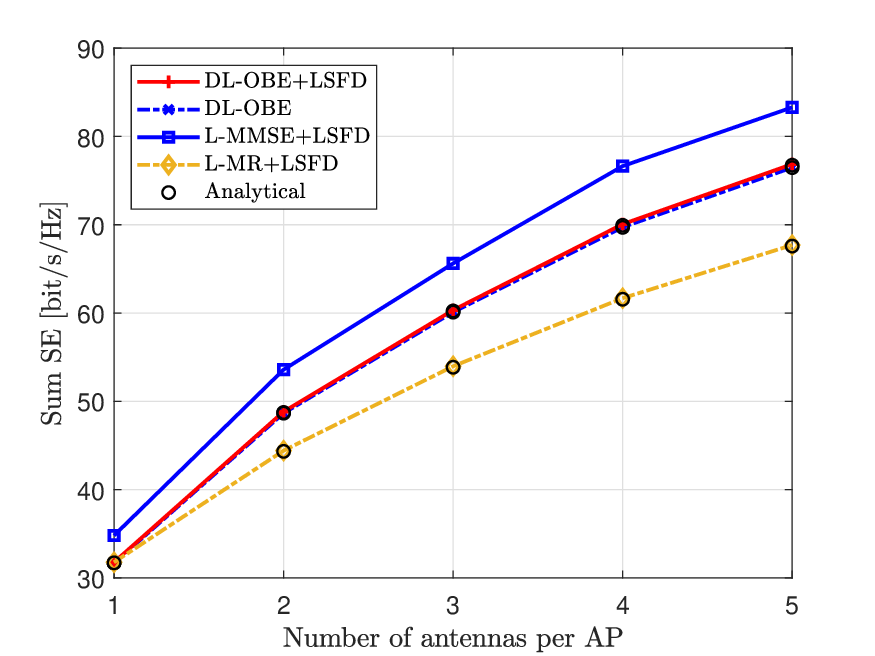}
\vspace{-0.37cm}
\caption{Sum SE performance against the number of antennas per AP for the EWDP and LSFD scheme with $M=40$, $K=20$, and $\tau_p=1$. \label{9}}
\vspace{-0.5cm}
\end{figure}

\begin{figure}[t]
\centering
\includegraphics[scale=0.45]{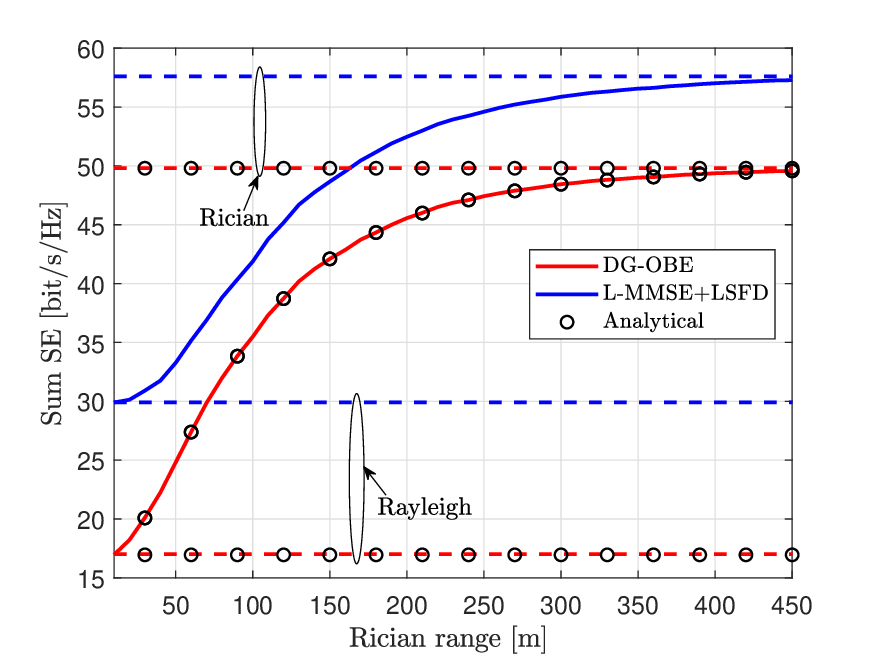}
\vspace{-0.37cm}
\caption{Sum SE performance versus different values of the ``Rician range" for the L-MMSE+LSFD and DG-OBE schemes with $M=20$, $K=20$, $N=4$, and $\tau_p=1$.\label{Rician_Region}}
\vspace{-0.5cm}
\end{figure}

\begin{figure}[t]
\centering
\includegraphics[scale=0.45]{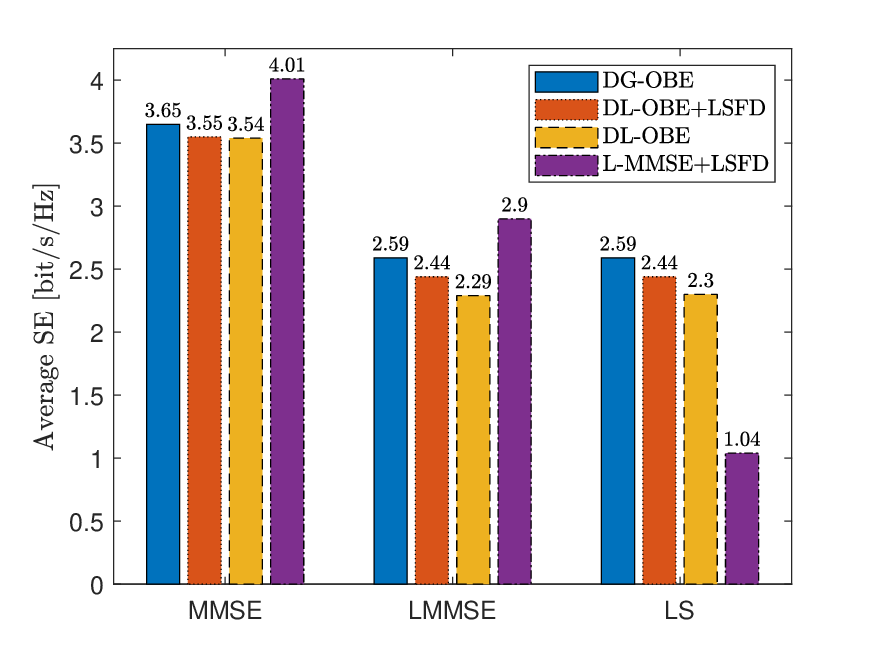}
\vspace{-0.37cm}
\caption{Average SE per UE for the distributed processing scheme over different channel estimators with $M=20$, $K=10$, $N=4$, and $\tau_p=1$.\label{Channel_Estimator}}
\vspace{-0.5cm}
\end{figure}

\begin{figure}[t]\centering
\vspace{0.3cm}
\subfigure[Centralized processing]{
\begin{minipage}{8cm}\centering
\includegraphics[scale=0.45]{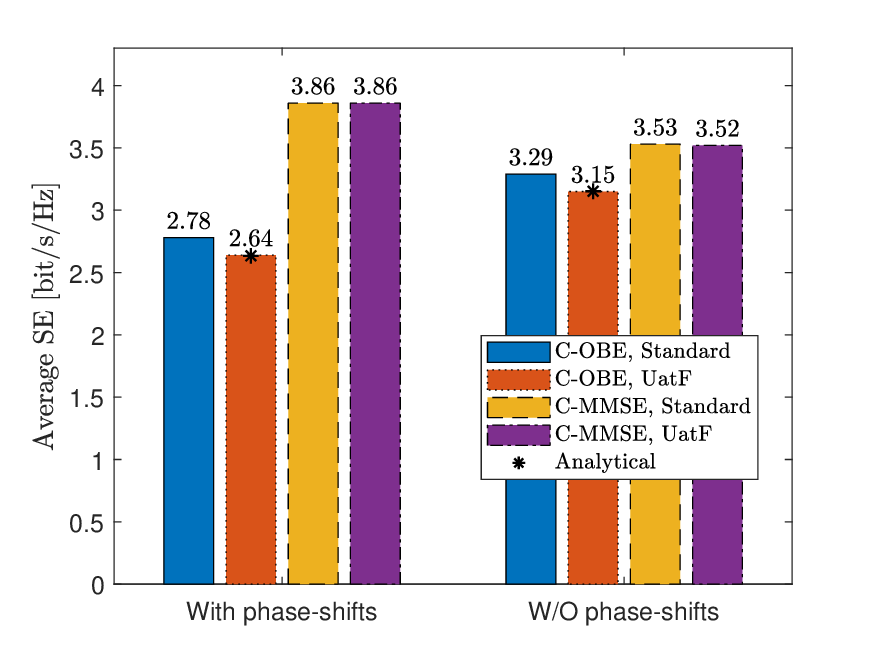}
\end{minipage}}
\subfigure[Distributed processing]{
\begin{minipage}{8cm}\centering
\includegraphics[scale=0.45]{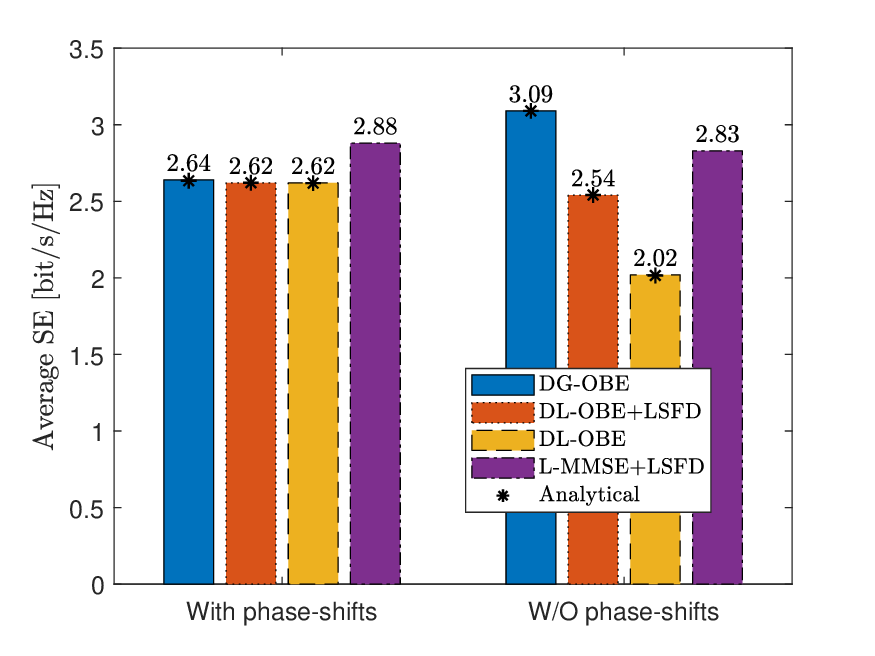}
\end{minipage}}
\vspace{-0.37cm}
\caption{Average SE per UE for different OBE schemes over the Rician fading channels with or without phase-shifts with $M=20$, $K=20$, $N=4$, and $\tau_p=1$.
\label{All_Schemes}}
\vspace{-0.8cm}
\end{figure}

To further study the SE performance in the scenarios with pilot contamination, we investigate the average SE per UE against the number of UEs in Fig.~\ref{4}. As observed, the C-OBE scheme performs steadily in the scenario with severe pilot contamination. There is only $3.8 \%$ performance gap between the SE with the C-OBE scheme with $\tau_p=1$ and $\tau_p=5$ for $K=30$. Moreover, the average SE performance gaps between the C-OBE and C-MMSE combining are $30.2 \%$ and $48.4 \%$ for $\tau_p=1$ and $\tau_p=5$ with $K=10$, respectively. These observations further demonstrate that the C-OBE combining is efficient to maintain robustness in scenarios with severe pilot contamination.

Next, we focus on the performance evaluation for the distributed processing schemes.\footnote{For convenience, we let the terminology ``equal weight distributed processing" and ``LSFD" denote the performance analysis frameworks introduced in \eqref{SINR_Distributed} and \eqref{optimal_lsfd_eq}, respectively.} Fig.~\ref{6} shows the average SE per UE against $M$ over the equal weight distributed processing (EWDP) and LSFD schemes. The L-MMSE+LSFD scheme\footnote{In the following, for convenience, we call ``A+LSFD" as the processing scheme, where ``A" combining scheme is applied at each AP and the optimal LSFD is implemented at the CPU. And ``A" denotes that ``A" combining scheme at each AP with no LSFD but just a simple equal weight scheme at the CPU.} achieves the best performance among all schemes. However, the DG-OBE is also efficient since the performance gap between DG-OBE and L-MMSE+LSFD is only $7.5 \%$ for $M=60$. Moreover, we find that the DG-OBE combining at each AP can achieve the same performance as the DG-OBE+LSFD scheme, which verifies the insights in Remark~\ref{noneed_LSFD}. This insightful observation demonstrates that the DG-OBE scheme can achieve an excellent SE performance with only the distributed combining applied at each AP and without the need to further implement the LSFD at the CPU. 

Fig.~\ref{8} shows the CDF curves of the SE per UE for the L-MMSE+LSFD and DG-OBE schemes with different values of $\tau_p$. As observed, the studied DG-OBE scheme is also efficient to maintain robustness in scenarios with severe pilot contamination, where the DG-OBE scheme can achieve approximately approaching SE performance for $\tau_p=1$ to that of $\tau_p=10$. Moreover, the average SE performance gaps between the DG-OBE and L-MMSE+LSFD schemes are $7.7 \%$ and $18.6 \%$ for $\tau_p=1$ and $\tau_p=10$, respectively, which also demonstrates that the DG-OBE scheme performs more efficiently in the scenarios with severe pilot contamination compared with the scenarios with little pilot contamination.

Next, we study the performance of the DL-OBE combining scheme. Fig.~\ref{9} shows the sum SE performance against $M$ under the EWDP and LSFD schemes. As observed, the DL-OBE scheme also performs well and there is only $9.1 \%$ performance gap between the DL-OBE and L-MMSE+LSFD schemes for $N=4$. More importantly, it is interesting to find that the application of the optimal LSFD strategy at the CPU just brings marginal SE performance improvement than that of the pure DL-OBE scheme without further LSFD strategy, e.g. only $0.5 \%$ SE improvement for the DL-OBE+LSFD scheme for $N=4$ compared with the DL-OBE scheme. Even though the DL-OBE scheme is locally designed based on local channel knowledge at each AP, it can achieve approximately the same performance as the DL-OBE+LSFD scheme, in which global statistical information is applied in the CPU to design the optimal LSFD strategy. This is because, due to the existence of random phase-shifts, many cross terms for different APs or different UEs become 0 and there is no further effective global information to be applied by the optimal LSFD strategy to enhance the performance. Besides, we notice that ``$\circ$" generated by the closed-form results in Th.~\ref{thm_LSFD_closed} and Th.~\ref{Dis_OBE_local_closed} match well with the curves generated by the Monte-Carlo simulation.

As explained earlier in this paper, all the OBE matrices are designed based on the channel statistics. In CF mMIMO systems, the semi-deterministic LoS components are dominant, which are widely utilized for the design of OBE matrices. We will now show the important effects of the LoS components on the achievable SE performance for the OBE schemes. In Fig.~\ref{Rician_Region}, we show the sum SE performance against the ``Rician range\footnote{If the distance for a particular AP-UE link falls within the Rician range, this link is modelled using the Rician fading model. Conversely, if the distance exceeds this range, this link is modelled using the Rayleigh fading model without a LoS component. Thus, increasing the value of Rician range increases the likelihood that a particular link will include a LoS component.}" for the L-MMSE+LSFD and DG-OBE schemes. We observe that the DG-OBE scheme performs more efficiently in the pure Rician scenario compared to the pure Rayleigh scenario. The performance gaps between the DG-OBE and L-MMSE+LSFD schemes are $13.5 \%$ and $43.1 \%$ over the Rician and Rayleigh scenarios, respectively. Moreover, as the Rician range increases, the achievable SE performance approaches to that of the pure Rician scenario. This is because continuously increasing Rician range can increase the number of Rician fading modelled links and thus enhances the system performance. We observe that ``$\circ$" generated by the closed-form results in Th.~\ref{thm_LSFD_closed} and Th.~\ref{Distributed_OBE_closed} match well with the curves generated by the Monte-Carlo simulation.

Fig.~\ref{Channel_Estimator} shows the average SE per UE for the distributed processing scheme when using different channel estimators. As observed, the distributed OBE schemes can also achieve excellent SE performance even without the prior knowledge of phase-shifts. Notably, the distributed OBE schemes can significantly outperform the L-MMSE combining-based scheme based on the LS estimator since the LS-based channel estimates show the poor ability to suppress interference. More interestingly, the distributed OBE schemes with the LS estimator can achieve similar SE performance compared to the distributed OBE schemes with the LMMSE estimator. This is because the distributed OBE schemes over both the LMMSE and LS estimators are linear to the $\mathbf{y}_{mk}^{p}$, and thus, can be uniformly constructed as $\mathbf{v}_{mk}=\mathbf{W}_{mk}\mathbf{y}_{mk}^{p}$. After optimizing the BE matrices $\mathbf{W}_{mk}$ based on different methodologies, the DG-OBE and DL-OBE schemes with the LS estimator are similar to those with the LMMSE estimator, which can achieve similar SE performance when using the LMMSE and LS estimators. Thus, under the scenario that the phase-shifts are not known or the scenario with the Rayleigh fading channel without the LoS component, if the OBE scheme is applied, only the low-complexity LS estimator is required to achieve similar SE performance as the computationally demanding MMSE-based channel estimators. Moreover, as observed, under the LMMSE and LS estimators, there is about $6.56 \%$ SE improvement, which is much larger than that of the MMSE estimator, for the DL-OBE+LSFD scheme compared to the DL-OBE scheme since phase-shifts are not known at each AP, and thus, the LSFD strategy can slightly enhance the SE performance based on global statistics.

To have a comprehensive overview of the proposed OBE schemes in this paper, we discuss the average SE per UE for various centralized and distributed processing schemes over both the Rician channel model with or without phase-shifts as shown in Fig.~\ref{All_Schemes}. The bar diagrams and numerical values are generated by the Monte-Carlo simulation. We observe in the centralized processing case in Fig.~\ref{All_Schemes} (a) that there is only $6.8\%$ average SE performance gap, measured by the standard capacity bound, between the C-OBE and the C-MMSE schemes in the scenario without phase-shifts, which is much smaller than the $28\%$ in the scenario with phase-shifts. Moreover, in the distributed processing case in Fig.~\ref{All_Schemes} (b), we observe that the DG-OBE scheme can outperform the L-MMSE+LSFD scheme in the scenario without phase-shifts. Besides, in the scenario without phase-shifts, the DL-OBE+LSFD scheme can achieve $25.7 \%$ average SE improvement compared to the DL-OBE scheme without the LSFD.  This occurs because, in the scenario without phase-shifts, numerous global cross terms from different APs or UEs can be effectively utilized by the LSFD strategy at the CPU to further enhance performance. However, in the scenario with random phase-shifts, most of these cross terms become $0$, significantly limiting the LSFD strategy's potential to improve performance. Thus, the DL-OBE scheme can achieve approximately the same performance as the DG-OBE and DL-OBE+LSFD schemes in the scenario with phase-shifts.

We will now extensively discuss all the OBE schemes proposed in this paper. Comparing Fig.~\ref{All_Schemes} (a) and Fig.~\ref{All_Schemes} (b), we find that, in the scenario with phase-shifts, the C-OBE scheme with the UatF bound achieves approximately the same performance as the distributed OBE schemes. As examined, in this scenario, the C-OBE matrix in \eqref{Centralized_obe_vector} is a block diagonal one, where $M$ matrices with the $N \times N$ dimension each are located at the diagonal. These $M$ matrices are approximately equal to those of DG-OBE matrices from all $M$ APs. Thus, this block diagonal form of the C-OBE matrix makes the achievable SE for the FCP scheme, measured by the UatF bound in \eqref{SE_centrlized_UatF}, approximately same as that of the DG-OBE scheme. In the scenario without phase-shifts, the performance gap between the C-OBE scheme under the UatF bound and the DG-OBE scheme is also marginal but the C-OBE scheme can achieve $6.5 \%$ SE improvement over the DG-OBE scheme with the aid of the standard bound. Finally, by summarizing the above important observations, we provide some promising insights for all three OBE schemes:

\textbf{\emph{(1) C-OBE is advocated in the scenario with limited fronthaul:}} As demonstrated in \cite[Fig. 6]{[162]},
the FCP scheme requires significantly less fronthaul signaling than the distributed processing scheme. The applications of both the DG-OBE and DL-OBE schemes necessitate extensive fronthaul signaling to transmit not only the soft data estimates but also channel statistics via fronthaul links. In contrast, the C-OBE scheme only requires the transmission of received pilot and data signals from all APs. Moreover, despite its higher computational complexity, the C-OBE scheme consistently achieves the best performance among all OBE schemes. Therefore, the C-OBE scheme is advocated in scenarios with limited fronthaul capacity, where all processing tasks can be centralized at the CPU.

\textbf{\emph{(2) DG-OBE is advocated in the scenario without phase-shifts in LoS components:}} When the fronthaul links are not limited, balancing the achievable performance and design complexity, the DG-OBE scheme is advocated in the scenarios without phase-shifts in LoS components. These scenarios include the scenario where the user is static and the scenario where the phase-shifts can be perfectly estimated and mitigated. In these scenarios, the DG-OBE scheme can achieve approaching performance to the C-OBE scheme and outperform the DL-OBE+LSFD scheme. To balance the trade-off between the achievable performance and design complexity, the DG-OBE scheme is most advocated in the scenario without phase-shifts.

\textbf{\emph{(3) DL-OBE is advocated in the scenario with phase-shifts in LoS components:}} In the scenarios with phase-shifts in LoS components, such as the mobile communication scenario and the scenario without perfect estimation and mitigation of phase-shifts, the DL-OBE scheme can approach the performance of the C-OBE and DG-OBE schemes with much lower computational complexity, thanks to the disappearance of most global cross terms caused by the random phase-shifts. Only local statistics information is required at each AP to design the DL-OBE matrix. Thus, the DL-OBE scheme is promising in the scenario with phase-shifts in LoS components.

\section{Conclusions}

We studied the OBE scheme design for CF mMIMO networks over Rician fading channels. The achievable SE performance analysis framework for the centralized and distributed processing schemes over the BE-structured combining was derived. We computed achievable SE expressions, measured by the UatF capacity bound, in novel closed-form. Then, one centralized and two distributed OBE schemes were proposed to maximize their respective achievable SE. OBE matrices in these OBE schemes were designed based on different levels of channel statistics. Notably, we obtained novel closed-form results for these OBE schemes. Numerical results verified our the matching accuracy between derived closed-form results and simulation results. It could be found that proposed OBE schemes could achieve excellent SE performance and exhibited robustness in scenarios with severe pilot contamination. More importantly, we observe that the C-OBE, DG-OBE, and DL-OBE schemes are advocated in the scenario with limited fronthaul, without phase-shifts in LoS components, and with phase-shifts in LoS components, respectively. 

\begin{appendices}
\vspace{-0.1cm}
\section{Useful Results}\label{useful}
In this appendix, some useful results are provided, which are applied for the derivations of results in this paper. For the vectors $\mathbf{x},\mathbf{y}\in \mathbb{C} ^N$ and the matrix $\mathbf{A}\in \mathbb{C} ^{N\times N}$, we have
\begin{equation}\label{trace}
\mathbf{x}^H\mathbf{A}\mathbf{y}=\mathrm{tr}(\mathbf{A}\mathbf{y}\mathbf{x}^H).
\end{equation}

For matrices $\mathbf{A}\in \mathbb{C} ^{N\times N}$, $\mathbf{B}\in \mathbb{C} ^{N\times N}$, $\mathbf{C}\in \mathbb{C} ^{N\times N}$, and $\mathbf{D}\in \mathbb{C} ^{N\times N}$ we have
\begin{align}
&\mathrm{vec}( \mathbf{ABC} ) =( \mathbf{C}^T\otimes \mathbf{A} ) \mathrm{vec}( \mathbf{B} ),  \label{vec_eq1}\\
&\mathrm{tr}( \mathbf{AB} ) =\mathrm{vec}( \mathbf{A}^H) ^H\mathrm{vec}( \mathbf{B} ),\label{vec_eq3}\\
&\mathrm{tr}( \mathbf{ABC} ) =\mathrm{vec}( \mathbf{A}^H ) ^H( \mathbf{I}_N\otimes \mathbf{B} ) \mathrm{vec}( \mathbf{C} ),\label{vec_eq2}\\
&\mathrm{tr}( \mathbf{A}^H\mathbf{BCD}^T ) =\mathrm{vec}( \mathbf{A} ) ^H( \mathbf{D}\otimes \mathbf{B} ) \mathrm{vec}( \mathbf{C} ). \label{vec_eq4}
\end{align}

\vspace*{-0.3cm}

\begin{figure*}[t]
{{\begin{align}\tag{38}\label{term1_expansion}
\mathbb{E} \{|\mathbf{v}_{k}^{H}\mathbf{g}_l|^2\}=\mathbb{E} \left\{ \mathbf{v}_{k}^{H}\mathbf{g}_l\mathbf{g}_{l}^{H}\mathbf{v}_k \right\} =\mathbb{E} \{( \underset{a}{\underbrace{\overline{\mathbf{g}}_{k}^{H}\mathbf{W}_{k}^{H}}}+\underset{b}{\underbrace{\widehat{\mathbf{g}}_{k,\mathrm{NLoS}}^{H}\mathbf{W}_{k}^{H}}} )\ ( \underset{c}{\underbrace{\overline{\mathbf{g}}_l}}+\underset{d}{\underbrace{\check{\mathbf{g}}_l}} ) ( \underset{e}{\underbrace{\overline{\mathbf{g}}_{l}^{H}}}+\underset{f}{\underbrace{\check{\mathbf{g}}_{l}^{H}}} ) ( \underset{g}{\underbrace{\mathbf{W}_k\overline{\mathbf{g}}_k}}+\underset{h}{\underbrace{\mathbf{W}_k\widehat{\mathbf{g}}_{k,\mathrm{NLoS}}}} )\}, 
\end{align}}
\hrulefill
\vspace*{-0.6cm}
}\end{figure*}

\begin{figure*}[t]
{{\begin{align}\tag{41}\label{Term1_closed}
&\mathbb{E} \{ \mathbf{v}_{k}^{H}\widehat{\mathbf{g}}_l\widehat{\mathbf{g}}_{l}^{H}\mathbf{v}_k \} =\mathrm{tr}( \mathbf{W}_{k}^{H}\overline{\mathbf{G}}_{ll}\mathbf{W}_k\overline{\mathbf{G}}_{kk} ) +p_l\tau _p\mathrm{tr}( \mathbf{W}_{k}^{H}\check{\mathbf{R}}_l\mathbf{\Psi }_{k}^{-1}\check{\mathbf{R}}_l\mathbf{W}_k\overline{\mathbf{G}}_{kk} ) +p_k\tau _p\mathrm{tr}( \mathbf{W}_{k}^{H}\overline{\mathbf{G}}_{ll}\mathbf{W}_k\check{\mathbf{R}}_k\mathbf{\Psi }_{k}^{-1}\check{\mathbf{R}}_k ) \\
&+p_kp_l\tau _{p}^{2}| \mathrm{tr}( \mathbf{W}_{k}^{H}\check{\mathbf{R}}_l\mathbf{\Psi }_{k}^{-1}\check{\mathbf{R}}_k ) |^2+p_kp_l\tau _{p}^{2}\mathrm{tr}( \mathbf{W}_{k}^{H}\check{\mathbf{R}}_l\mathbf{\Psi }_{k}^{-1}\check{\mathbf{R}}_l\mathbf{W}_k\check{\mathbf{R}}_k\mathbf{\Psi }_{k}^{-1}\check{\mathbf{R}}_k ) \notag
\end{align}}
\hrulefill
\vspace*{-0.6cm}
}\end{figure*}

\section{Proof of Theorem~\ref{thm_centralized_closed}}\label{app_cenralized_closed}

In this appendix, we provide detailed proof steps of Theorem~\ref{thm_centralized_closed}. We begin from the numerator of \eqref{SINR_centrlized_UatF}. We have
\begin{equation}
\begin{aligned}
&\mathbb{E} \{ \mathbf{v}_{k}^{H}\mathbf{g}_k \} =\mathbb{E} \{ \widehat{\mathbf{g}}_{k}^{H}\mathbf{W}_{k}^{H}\mathbf{g}_k \} \overset{( a )}{=}\mathrm{tr}( \mathbf{W}_{k}^{H}\mathbb{E} \{ \mathbf{g}_k\widehat{\mathbf{g}}_{k}^{H} \} ) \\
&\overset{( b )}{=}\mathrm{tr}( \mathbf{W}_{k}^{H}\mathbb{E} \{ \widehat{\mathbf{g}}_k\widehat{\mathbf{g}}_{k}^{H} \} ) =\mathrm{tr}( \mathbf{W}_{k}^{H}\overline{\mathbf{R}}_k ), 
\end{aligned}
\end{equation}
where step (a) follows from \eqref{trace} and step (b) is based on the independent feature between the $\tilde{\mathbf{g}}_k$ and $\widehat{\mathbf{g}}_k$ and zero-mean feature of $\tilde{\mathbf{g}}_k$. Besides, we define $\overline{\mathbf{R}}_k\triangleq \mathbb{E} \{ \widehat{\mathbf{g}}_k\widehat{\mathbf{g}}_{k}^{H} \}
=\overline{\mathbf{G}}_{kk}+p_k\tau _p\check{\mathbf{R}}_k\mathbf{\Psi }_{k}^{-1}\check{\mathbf{R}}_k$, where $\overline{\mathbf{G}}_{kk}=\mathbb{E} \{ \overline{\mathbf{g}}_k\overline{\mathbf{g}}_{k}^{H} \} =\mathrm{diag}[ \overline{\mathbf{G}}_{1kk},\dots ,\overline{\mathbf{G}}_{Mkk} ] $
due to the independent characteristics of $\{\varphi_{mk}\}$ for different AP-UE pairs, that is $\mathbb{E} \{ e^{j\theta _{mk}}e^{-j\theta _{nl}} \}=\mathbb{E} \{ e^{j\theta _{mk}} \}\mathbb{E} \{e^{-j\theta _{nl}} \}=0$ if $m\ne n$ or $k\ne l$. Meanwhile, it can be easily verified that  $\mathbb{E} \{ e^{j\theta _{mk}}\}=0$ according to the random distribution of $\theta _{mk}\sim \mathcal{U} [ -\pi ,\pi ]$.

\begin{figure*}[t]
{{\begin{align}\tag{44}\label{SINR_centralized_expansion}
\mathrm{SINR}_{k}^{\mathrm{c},\mathrm{UatF}}=\frac{p_k| \mathrm{tr}( \mathbf{W}_{k}^{H}\mathbb{E} \{ \mathbf{g}_k\widehat{\mathbf{g}}_{k}^{H} \} ) |^2}{\sum_{l=1}^K{p_l\mathbb{E} \{ \mathrm{tr}( \mathbf{W}_{k}^{H}\mathbf{g}_l\mathbf{g}_{l}^{H}\mathbf{W}_k\widehat{\mathbf{g}}_k\widehat{\mathbf{g}}_{k}^{H} ) \}}-p_k| \mathrm{tr}( \mathbf{W}_{k}^{H}\mathbb{E} \{ \mathbf{g}_k\widehat{\mathbf{g}}_{k}^{H} \} ) |^2+\sigma ^2\mathrm{tr}( \mathbf{W}_{k}^{H}\mathbf{W}_k\mathbb{E} \{ \widehat{\mathbf{g}}_k\widehat{\mathbf{g}}_{k}^{H} \} )}
\end{align}}
\hrulefill
\vspace*{-0.7cm}
}\end{figure*}

\begin{figure*}[t]
{{\begin{align}\tag{45}\label{SINR_centralized_vec}
\mathrm{SINR}_{k}^{\mathrm{c},\mathrm{UatF}}=\frac{p_k|\mathbf{w}_{k}^{H}\mathrm{vec(}\mathbb{E} \{\mathbf{g}_k\widehat{\mathbf{g}}_{k}^{H}\})|^2}{\mathbf{w}_{k}^{H}\left[ \sum_{l=1}^K{p_l\mathbb{E} \{[(\widehat{\mathbf{g}}_k\widehat{\mathbf{g}}_{k}^{H})^T\otimes (\mathbf{g}_l\mathbf{g}_{l}^{H})]\}-p_k|\mathbf{w}_{k}^{H}\mathrm{vec(}\mathbb{E} \{\mathbf{g}_k\widehat{\mathbf{g}}_{k}^{H}\})|^2+(\mathbb{E} \{\widehat{\mathbf{g}}_k\widehat{\mathbf{g}}_{k}^{H}\}^T\otimes \mathbf{I}_{MN})} \right] \mathbf{w}_k}
\end{align}}
\hrulefill
\vspace*{-0.7cm}
}\end{figure*}

\begin{figure*}[t]
{{\begin{align}\tag{49}\label{SINR_Distributed_Closed}
\overline{\widetilde{\mathrm{SINR}}}_{k}^{\mathrm{LSFD}}=\frac{p_k| \sum_{m=1}^M{\mathrm{tr}( \mathbf{W}_{mk}^{H}\overline{\mathbf{R}}_{mk} )} |^2}{\left( \begin{array}{c}
	\sum_{l=1}^K{\sum_{m=1}^M{p_l\varphi _{mkl}}}+\sum_{l\in \mathcal{P} _k}{\sum_{m=1}^M{p_l\upsilon _{mkl}}}+\sum_{l\in \mathcal{P} _k}{p_l( | \sum_{m=1}^M{\gamma _{mkl}} |^2-\sum_{m=1}^M{| \gamma _{mkl} |^2} )}\\
	-p_k| \sum_{m=1}^M{\mathrm{tr}( \mathbf{W}_{mk}^{H}\overline{\mathbf{R}}_{mk} )} |^2+\sigma ^2\sum_{m=1}^M{\mathrm{tr}( \mathbf{W}_{mk}\overline{\mathbf{R}}_{mk} \mathbf{W}_{mk}^{H}) }\\
\end{array} \right)}
\end{align}}
\hrulefill
\vspace*{-0.7cm}
}\end{figure*}

As for $\mathbb{E} \{ | \mathbf{v}_{k}^{H}\mathbf{g}_l |^2 \}$, if $l\notin \mathcal{P} _k$, $\widehat{\mathbf{g}}_k$ and $\mathbf{g}_l$ are independent. We denote  $\mathbb{E} \{ | \mathbf{v}_{k}^{H}\mathbf{g}_l |^2 \}$ in \eqref{term1_expansion} on the top of the next page, where $\widehat{\mathbf{g}}_{k,\mathrm{NLoS}}=\sqrt{p_k}\check{\mathbf{R}}_k\mathbf{\Psi }_{k}^{-1}( \mathbf{y}_{k}^{p}-\overline{\mathbf{y}}_{k}^{p} )$. Then, we can compute terms in \eqref{term1_expansion} in closed-form by expanding them as $\mathbb{E} \left\{ aceg \right\}  =\mathrm{tr}( \mathbf{W}_{k}^{H}\overline{\mathbf{G}}_{ll}\mathbf{W}_k\overline{\mathbf{G}}_{kk} )$, $\mathbb{E} \{ adfg \} =\mathrm{tr}( \mathbf{W}_{k}^{H}\check{\mathbf{R}}_l\mathbf{W}_k\overline{\mathbf{G}}_{kk} ) $, $\mathbb{E} \{ bceh \}  =\mathrm{tr}( \mathbf{W}_{k}^{H}\overline{\mathbf{G}}_{ll}\mathbf{W}_k \widehat{\mathbf{R}}_k) $, and $\mathbb{E} \{ bdfh \}  =\mathrm{tr}( \mathbf{W}_{k}^{H}\mathbb{E} \{ \check{\mathbf{g}}_l\check{\mathbf{g}}_{l}^{H} \} \mathbf{W}_k\mathbb{E} \{ \widehat{\mathbf{g}}_{k,\mathrm{NLoS}}\widehat{\mathbf{g}}_{k,\mathrm{NLoS}}^{H} \} ) =\mathrm{tr}( \mathbf{W}_{k}^{H}\check{\mathbf{R}}_l\mathbf{W}_k\widehat{\mathbf{R}}_k ) $. The other terms are all zero, which can be easily verified. In summary, for $l\notin \mathcal{P} _k$, we have 
$\mathbb{E} \{ \mathbf{v}_{k}^{H}\mathbf{g}_l\mathbf{g}_{l}^{H}\mathbf{v}_k \} =\mu _{kl}$.
If $l\in \mathcal{P} _k\backslash \{ k \}$, $\widehat{\mathbf{g}}_k$ and $\mathbf{g}_l$ are not independent since $\mathbf{y}_{k}^{p}$ contains $\mathbf{g}_l$. By following the method in \cite[Appendix D]{ozdogan2019massive}, we can construct $\widehat{\mathbf{g}}_k$ and $\widehat{\mathbf{g}}_l$ as $\widehat{\mathbf{g}}_k=\overline{\mathbf{g}}_k+\sqrt{p_k\tau _p}\check{\mathbf{R}}_k\mathbf{\Psi }_{k}^{-\frac{1}{2}}\mathbf{q}$ and $\widehat{\mathbf{g}}_l =\overline{\mathbf{g}}_l+\sqrt{p_l\tau _p}\check{\mathbf{R}}_l\mathbf{\Psi }_{k}^{-\frac{1}{2}}\mathbf{q}$, respectively, where $\mathbf{q}\sim \mathcal{N} _{\mathbb{C}}( \mathbf{0},\mathbf{I}_{MN} ) $. We formulate $\mathbb{E} \{ | \mathbf{v}_{k}^{H}\mathbf{g}_l |^2 \}$ as 
\setcounter{equation}{38}
\begin{equation}\label{term1_fenli}
\mathbb{E} \{ | \mathbf{v}_{k}^{H}\mathbf{g}_l |^2 \}=\mathbb{E} \{ \mathbf{v}_{k}^{H}\widehat{\mathbf{g}}_l\widehat{\mathbf{g}}_{l}^{H}\mathbf{v}_k \} +\mathbb{E} \{ \mathbf{v}_{k}^{H}\tilde{\mathbf{g}}_l\tilde{\mathbf{g}}_{l}^{H}\mathbf{v}_k \}. 
\end{equation}
For $\mathbb{E} \{ \mathbf{v}_{k}^{H}\widehat{\mathbf{g}}_l\widehat{\mathbf{g}}_{l}^{H}\mathbf{v}_k \}$, we have
\begin{equation}\label{term2_expansion}
\begin{aligned}
&\mathbb{E} \{ \mathbf{v}_{k}^{H}\widehat{\mathbf{g}}_l\widehat{\mathbf{g}}_{l}^{H}\mathbf{v}_k \} =\mathbb{E} \{ ( \underset{a}{\underbrace{\overline{\mathbf{g}}_{k}^{H}\mathbf{W}_{k}^{H}}}+\underset{b}{\underbrace{\sqrt{p_k\tau _p}\mathbf{q}^H\mathbf{\Psi }_{k}^{-\frac{1}{2}}\check{\mathbf{R}}_k\mathbf{W}_{k}^{H}}} ) \\
&( \underset{c}{\underbrace{\overline{\mathbf{g}}_l}}+\underset{d}{\underbrace{\sqrt{p_l\tau _p}\check{\mathbf{R}}_l\mathbf{\Psi }_{k}^{-\frac{1}{2}}\mathbf{q}}} ) ( \underset{e}{\underbrace{\overline{\mathbf{g}}_{l}^{H}}}+\underset{f}{\underbrace{\sqrt{p_l\tau _p}\mathbf{q}^H\mathbf{\Psi }_{k}^{-\frac{1}{2}}\check{\mathbf{R}}_l}} )\\
&( \underset{g}{\underbrace{\mathbf{W}_k\overline{\mathbf{g}}_k}}+\underset{h}{\underbrace{\sqrt{p_k\tau _p}\mathbf{W}_k\check{\mathbf{R}}_k\mathbf{\Psi }_{k}^{-\frac{1}{2}}\mathbf{q}}} ) \}.
\end{aligned}
\end{equation}
By expanding terms in \eqref{term2_expansion}, we can compute them in closed-form as 
$\mathbb{E} \{ aceg \} =\mathrm{tr}( \mathbf{W}_{k}^{H}\overline{\mathbf{G}}_{ll}\mathbf{W}_k\overline{\mathbf{G}}_{kk} ) $, $\mathbb{E} \{ adfg \} =p_l\tau _p\mathrm{tr}( \mathbf{W}_{k}^{H}\check{\mathbf{R}}_l\mathbf{\Psi }_{k}^{-1}\check{\mathbf{R}}_l\mathbf{W}_k\overline{\mathbf{G}}_{kk} ) $, 
$\mathbb{E} \left\{ bceh \right\}=p_k\tau _p\mathrm{tr}( \mathbf{W}_{k}^{H}\overline{\mathbf{G}}_{ll}\mathbf{W}_k\check{\mathbf{R}}_k\mathbf{\Psi }_{k}^{-1}\check{\mathbf{R}}_k) $, and 
$\mathbb{E} \{ bdfh \} \overset{( a )}{=}p_l\tau _{p}\mathrm{tr}( \mathbf{W}_{k}^{H}\check{\mathbf{R}}_l\mathbf{\Psi }_{k}^{-1}\check{\mathbf{R}}_l\mathbf{W}_k\widehat{\mathbf{R}}_k  )+p_kp_l\tau _{p}^{2}| \mathrm{tr}( \mathbf{W}_{k}^{H}\check{\mathbf{R}}_l\mathbf{\Psi }_{k}^{-1}\check{\mathbf{R}}_k ) |^2$
where step (a) is according to \cite[Eq. 80]{ozdogan2019massive} by letting $( \mathbf{R}_{x}^{H} ) ^{\frac{1}{2}}=\mathbf{\Psi }_{k}^{-\frac{1}{2}}\check{\mathbf{R}}_k$, $\mathbf{R}_{y}^{{\frac{1}{2}}}=\check{\mathbf{R}}_l\mathbf{\Psi }_{k}^{-\frac{1}{2}}$, and $\mathbf{B}=\mathbf{W}_{k}^{H}$, respectively. The other terms are zero due to the circular symmetry characteristic of $\mathbf{q}$ and the independent characteristic of $\{\theta_{mk}\}$ for different AP-UE pairs.
In summary, for $l\in \mathcal{P} _k\backslash \{ k \}$, we can compute $\mathbb{E} \{ \mathbf{v}_{k}^{H}\widehat{\mathbf{g}}_l\widehat{\mathbf{g}}_{l}^{H}\mathbf{v}_k \}$ in closed-form in \eqref{Term1_closed} on the top of the next page. As for $\mathbb{E} \{ \mathbf{v}_{k}^{H}\tilde{\mathbf{g}}_l\tilde{\mathbf{g}}_{l}^{H}\mathbf{v}_k \} $, relying on the independence feature between $\widehat{\mathbf{g}}_k$ and $\tilde{\mathbf{g}}_{l}$, we can compute it in closed form
\setcounter{equation}{41}
\begin{equation}\label{error_closed}
\begin{aligned}
&\mathbb{E} \{ \mathbf{v}_{k}^{H}\tilde{\mathbf{g}}_l\tilde{\mathbf{g}}_{l}^{H}\mathbf{v}_k \} =\mathbb{E} \{ | ( \mathbf{W}_k\widehat{\mathbf{g}}_k ) ^H\tilde{\mathbf{g}}_l |^2 \} \\
&=p_k\tau _p\mathrm{tr}( \mathbf{W}_{k}^{H}\mathbf{C}_l\mathbf{W}_k\check{\mathbf{R}}_k\mathbf{\Psi }_{k}^{-1}\check{\mathbf{R}}_k ) +\mathrm{tr}( \mathbf{W}_{k}^{H}\mathbf{C}_l\mathbf{W}_k\overline{\mathbf{G}}_{kk} ) 
\end{aligned}
\end{equation}
by applying \cite[Lemma 4]{ozdogan2019massive} with $\mathbf{R}_x=p_k\tau _p\check{\mathbf{R}}_k\mathbf{\Psi }_{k}^{-1}\check{\mathbf{R}}_k$,
$\mathbf{R}_y=\mathbf{C}_l$, and $\mathbf{B}=\mathbf{W}_{k}^{H}$, respectively. By plugging \eqref{Term1_closed} and \eqref{error_closed} into \eqref{term1_fenli}, we can compute $\mathbb{E} \left\{ \mathbf{v}_{k}^{H}\mathbf{g}_l\mathbf{g}_{l}^{H}\mathbf{v}_k \right\}$ in closed-form as $\mu _{kl}+\varepsilon _{kl}$ by applying the fact that $p_l\tau _p\check{\mathbf{R}}_l\mathbf{\Psi }_{k}^{-1}\check{\mathbf{R}}_l=\check{\mathbf{R}}_l-\mathbf{C}_l$. For $l=k$, besides terms $\mu _{kl}+\varepsilon _{kl}$ for $l\in \mathcal{P} _k\backslash \{ k \}$, there are also additional terms $\mathbb{E} \{acfh\}=p_k\tau _p\mathrm{tr(}\mathbf{W}_{k}^{H}\overline{\mathbf{G}}_{kk})\mathrm{tr(}\mathbf{W}_k\check{\mathbf{R}}_k\mathbf{\Psi }_{k}^{-1}\check{\mathbf{R}}_k)$ and $\mathbb{E} \{bdeg\}=\!p_k\tau _p\mathrm{tr(}\mathbf{W}_{k}^{H}\check{\mathbf{R}}_k\mathbf{\Psi }_{k}^{-1}\check{\mathbf{R}}_k)\mathrm{tr(}\mathbf{W}_k\overline{\mathbf{G}}_{kk})$. Moreover, the computation of $\mathbb{E} \{ aceg \} =\mathbb{E} \{ \overline{\mathbf{g}}_{k}^{H}\mathbf{W}_{k}^{H}\overline{\mathbf{g}}_k\overline{\mathbf{g}}_{k}^{H}\mathbf{W}_k\overline{\mathbf{g}}_k \} =\mathbb{E} \{ | \overline{\mathbf{g}}_{k}^{H}\mathbf{W}_{k}^{H}\overline{\mathbf{g}}_k |^2 \} $ is quite tricky since there are $M$ different phase-shifts in $\overline{\mathbf{g}}_k$ from $M$ APs. By constructing $\mathbf{W}_{k}$ into the block matrix form, we can decompose $\mathbb{E} \{ | \overline{\mathbf{g}}_{k}^{H}\mathbf{W}_{k}^{H}\overline{\mathbf{g}}_k |^2 \}$ as 
\begin{equation}\label{LoS_expansion}
\begin{aligned}
&\mathbb{E} \{|\overline{\mathbf{g}}_{k}^{H}\mathbf{W}_{k}^{H}\overline{\mathbf{g}}_k|^2\}\!\!=\!\!\sum_{\mathbf{m}}^{}{\mathbb{E} \{\overline{\mathbf{g}}_{m_1k}^{H}e^{-j\theta _{m_1}k}\mathbf{W}_{k,m_2m_1}^{H}\overline{\mathbf{g}}_{m_2k}e^{j\theta _{m_2}k}}\\
&\overline{\mathbf{g}}_{m_3k}^{H}e^{-j\theta _{m_3}k}\mathbf{W}_{k,m_3m_4}\overline{\mathbf{g}}_{m_4k}e^{j\theta _{m_4}k}\},
\end{aligned}
\end{equation}
where $\sum_{\mathbf{m}}^{}{}=\sum_{m_1=1}^M{\sum_{m_2=1}^M{\sum_{m_3=1}^M{\sum_{m_4=1}^M{}}}}$. Note that different phase-shifts with different $m$-subscript are i.i.d. and many cross-terms equal to 0 due to the existence of phase-shifts. By discussing all possible combinations of $m_1\sim m_4$, we 
can compute \eqref{LoS_expansion} into closed-form as shown in \eqref{LoS_Closed}.
As for $\mathbb{E} \{ \| \mathbf{v}_k \| ^2 \}$, we have
$\mathbb{E} \{ \| \mathbf{v}_k \| ^2 \} =\mathbb{E} \{ \widehat{\mathbf{g}}_{k}^{H}\mathbf{W}_{k}^{H}\mathbf{W}_k\widehat{\mathbf{g}}_k \}=\mathrm{tr}( \mathbf{W}_{k}^{H}\mathbf{W}_k\mathbb{E} \{ \widehat{\mathbf{g}}_k\widehat{\mathbf{g}}_{k}^{H} \} ) =\mathrm{tr}( \mathbf{W}_{k}^{H}\mathbf{W}_k\overline{\mathbf{R}}_k ).
$
Finally, combining all results in this appendix, we finish the proof. 

\section{Proof of Theorem~\ref{thm_LSFD_closed}}\label{closed-LSFD-app}
\vspace*{-0.2cm}
We provide detailed proof steps of Theorem~\ref{thm_LSFD_closed} in this appendix. We start from $\mathbb{E} \left\{ \mathbf{b}_{kk} \right\}$ and we can easily compute the $m$-th element of it in closed-form as $\mathbb{E} \{ \mathbf{v}_{mk}^{H}\mathbf{g}_{mk} \} =\mathrm{tr}( \mathbf{W}_{mk}^{H}\overline{\mathbf{R}}_{mk} ) $ based on the results in Appendix~\ref{app_cenralized_closed}. As for $[ \mathbf{\Xi }_{kl} ] _{nm}=\mathbb{E} \{( \mathbf{v}_{mk}^{H}\mathbf{g}_{ml} ) ^H( \mathbf{v}_{nk}^{H}\mathbf{g}_{nl} ) \}$, we can compute it in closed-form for all possible AP-UE combinations. For $m\ne n, l\notin \mathcal{P} _k$, we have $\mathbb{E} \{ ( \mathbf{v}_{mk}^{H}\mathbf{g}_{ml} ) ^H( \mathbf{v}_{nk}^{H}\mathbf{g}_{nl} ) \} =\mathbb{E} \{ \mathbf{g}_{ml}^{H}\mathbf{v}_{mk} \} \mathbb{E} \{ \mathbf{v}_{nk}^{H}\mathbf{g}_{nl} \} =0$. For $m\ne n,l\in \mathcal{P} _k$, $\mathbb{E} \{ ( \mathbf{v}_{mk}^{H}\mathbf{g}_{ml} ) ^H( \mathbf{v}_{nk}^{H}\mathbf{g}_{nl} ) \} =\mathrm{tr}( \mathbf{W}_{mk}\mathbf{B}_{mkl} ) \mathrm{tr}( \mathbf{W}_{nk}^{H}\mathbf{B}_{nlk} )$, where 
$\mathbf{B}_{mkl}=\mathbb{E} \{ \widehat{\mathbf{g}}_{mk}\mathbf{g}_{ml}^{H} \} =\sqrt{p_kp_l}\tau _p\check{\mathbf{R}}_{mk}\mathbf{\Psi }_{mk}^{-1}\check{\mathbf{R}}_{ml}$ for $l\in \mathcal{P} _k\backslash \{ k \}$ and $\overline{\mathbf{G}}_{mkk}+\sqrt{p_kp_l}\tau _p\check{\mathbf{R}}_{mk}\mathbf{\Psi }_{mk}^{-1}\check{\mathbf{R}}_{ml}$ for $l=k$. For $m=n,l\notin \mathcal{P} _k$, $\widehat{\mathbf{g}}_{mk}$ and $\mathbf{g}_{ml}$ are independent, so we can derive 
$\mathbb{E} \{ ( \mathbf{v}_{mk}^{H}\mathbf{g}_{ml} ) ^H( \mathbf{v}_{nk}^{H}\mathbf{g}_{nl} ) \} =\mathrm{tr}( \mathbf{W}_{mk}^{H}\overline{\mathbf{G}}_{mll}\mathbf{W}_{mk}\overline{\mathbf{G}}_{mkk} ) +\mathrm{tr}( \mathbf{W}_{mk}^{H}\check{\mathbf{R}}_{ml}\mathbf{W}_{mk}\overline{\mathbf{G}}_{mkk} ) +\mathrm{tr}( \mathbf{W}_{mk}^{H}\overline{\mathbf{G}}_{mll}\widehat{\mathbf{R}}_{mk}\mathbf{W}_{mk} ) +\mathrm{tr}( \mathbf{W}_{mk}^{H}\check{\mathbf{R}}_{ml}\mathbf{W}_{mk}\widehat{\mathbf{R}}_{mk} )$,
by the method discussed in \eqref{term1_expansion}. As for $m=n,l\in \mathcal{P} _k$, $\widehat{\mathbf{g}}_{mk}$ and $\mathbf{g}_{ml}$ are not independent. Based on the similar method as \eqref{term1_fenli}, we can easily compute $\mathbb{E} \{ ( \mathbf{v}_{mk}^{H}\mathbf{g}_{ml} ) ^H( \mathbf{v}_{nk}^{H}\mathbf{g}_{nl} ) \} =\varphi _{mkl}+\upsilon _{mkl}$. Moreover, we have $\mathbb{E} \{ \| \mathbf{v}_{mk} \| ^2 \}=\mathrm{tr}( \mathbf{W}_{mk}^{H}\mathbf{W}_{mk}\overline{\mathbf{R}}_{mk} )$. Based on all derived results, we can easily derive the closed-form results in Theorem~\ref{thm_LSFD_closed}.

\vspace{-0.5cm}
\section{Proof of Corollary~\ref{C_OBE_UatF}}\label{app_centralized_obe}
\vspace*{-0.25cm}
By applying \eqref{trace}, we can represent $\mathrm{SINR}_{k}^{\mathrm{c},\mathrm{UatF}}$ in \eqref{SINR_OBE} as \eqref{SINR_centralized_expansion} on the top of this page. By vectorizing $\mathbf{w}_k=\mathrm{vec}( \mathbf{W}_k ) \in \mathbb{C} ^{M^2N^2}$, we can derive $\mathrm{tr}( \mathbf{W}_{k}^{H}\mathbb{E} \{ \mathbf{g}_k\widehat{\mathbf{g}}_{k}^{H} \} ) =\mathbf{w}_{k}^{H}\mathrm{vec}( \mathbb{E} \{ \mathbf{g}_k\widehat{\mathbf{g}}_{k}^{H} \} ) $, applying \eqref{vec_eq3} by letting $\mathbf{A}=\mathbf{W}_{k}$ and $\mathbf{B}=\mathbb{E} \{ \mathbf{g}_k\widehat{\mathbf{g}}_{k}^{H} \}$, respectively. As for $\mathbb{E} \{ \mathrm{tr}( \mathbf{W}_{k}^{H}\mathbf{g}_l\mathbf{g}_{l}^{H}\mathbf{W}_k\widehat{\mathbf{g}}_k\widehat{\mathbf{g}}_{k}^{H} ) \}$, we can apply \eqref{vec_eq4} to derive
$
\mathbb{E} \{ \mathrm{tr}( \mathbf{W}_{k}^{H}\mathbf{g}_l\mathbf{g}_{l}^{H}\mathbf{W}_k\widehat{\mathbf{g}}_k\widehat{\mathbf{g}}_{k}^{H} ) \} \!\!=\!\!\mathbf{w}_{k}^{H}\mathbb{E} \{ [ ( \widehat{\mathbf{g}}_k\widehat{\mathbf{g}}_{k}^{H} ) ^T\otimes ( \mathbf{g}_l\mathbf{g}_{l}^{H} ) ] \} \mathbf{w}_k,
$
by letting $\mathbf{A}=\mathbf{W}_{k}$, $\mathbf{B}=\mathbf{g}_l\mathbf{g}_{l}^{H}$, $\mathbf{C}=\mathbf{W}_{k}$, and $\mathbf{D}^T=\widehat{\mathbf{g}}_k\widehat{\mathbf{g}}_{k}^{H} $, respectively. For $\mathrm{tr}( \mathbf{W}_{k}^{H}\mathbf{W}_k\mathbb{E} \{ \widehat{\mathbf{g}}_k\widehat{\mathbf{g}}_{k}^{H} \} )$, based on \eqref{vec_eq2}, we can obtain
$\mathrm{tr}( \mathbf{W}_{k}^{H}\mathbf{W}_k\mathbb{E} \{ \widehat{\mathbf{g}}_k\widehat{\mathbf{g}}_{k}^{H} \} )=\mathbf{w}_{k}^{H}( \mathbb{E} \{ \widehat{\mathbf{g}}_k\widehat{\mathbf{g}}_{k}^{H} \} ^T\otimes \mathbf{I}_{MN} ) \mathbf{w}_k$, by letting $\mathbf{A}=\mathbf{W}_{k}$, $\mathbf{B}=\mathbf{I}_{MN}$, $\mathbf{C}=\mathbf{W}_{k}$, and $\mathbf{D}^T=\mathbb{E} \{ \widehat{\mathbf{g}}_k\widehat{\mathbf{g}}_{k}^{H} \} $, respectively. Thus, by combining all these terms, we can construct \eqref{SINR_centralized_expansion} in \eqref{SINR_centralized_vec} on the top of this page.
Note that \eqref{SINR_centralized_vec} is a standard Rayleigh quotient with respect to $\mathbf{w}_k$. Thus, by applying \cite[Lemma B.10]{8187178}, we can derive the optimal $ \mathbf{w}_{k}^{*}$, which can maximize \eqref{SINR_centralized_vec}, in \eqref{Optimal_centralized_obe}, leading to the maximum SINR value in \eqref{C_SINR_Max}.
\vspace{-0.3cm}
\section{Proof of Theorem~\ref{Centralized_OBE_closed} }\label{app_centralized_obe_closed}
We first reformulate terms of $\overline{\mathrm{SINR}}_{k}^{\mathrm{c},\mathrm{UatF}}$ given in \eqref{Centrlized_closed}. By applying \eqref{vec_eq3}, we have $\mathrm{tr}( \mathbf{W}_{k}^{H}\overline{\mathbf{R}}_k ) =\mathbf{w}_{k}^{H}\overline{\mathbf{r}}_k$. Moreover, by applying results in Appendix~\ref{useful}, we can easily represent $\mu _{kl}$, $\varepsilon _{kl}$, and $\mathrm{tr}( \mathbf{W}_k\overline{\mathbf{R}}_k \mathbf{W}_{k}^{H})$ in \eqref{Centrlized_closed} as $
\mu_{kl}=\mathbf{w}_{k}^{H}( \overline{\mathbf{G}}_{kk}^{T}\otimes \overline{\mathbf{G}}_{ll} ) \mathbf{w}_k+\mathbf{w}_{k}^{H}( \overline{\mathbf{G}}_{kk}^{T}\otimes \check{\mathbf{R}}_l ) \mathbf{w}_k+\mathbf{w}_{k}^{H}( \widehat{\mathbf{R}}_{k}^{T}\otimes \overline{\mathbf{G}}_{ll} ) \mathbf{w}_k+\mathbf{w}_{k}^{H}( \widehat{\mathbf{R}}_{k}^{T}\otimes \check{\mathbf{R}}_l ) \mathbf{w}_k$, $\varepsilon _{kl}=p_kp_l\tau _{p}^{2}| \mathbf{w}_{k}^{H}\tilde{\mathbf{r}}_{lk} |^2$, and $\mathrm{tr}( \mathbf{W}_k\overline{\mathbf{R}}_k \mathbf{W}_{k}^{H}) =\mathbf{w}_{k}^{H}( \overline{\mathbf{R}}_{k}^{T}\otimes \mathbf{I}_{MN} ) \mathbf{w}_k$, respectively. 
As for $\omega _k$ in \eqref{LoS_Closed}, the first three terms can also be easily reformulated as $p_k\tau _p\mathbf{w}_{k}^{H}\overline{\mathbf{g}}_{kk}\tilde{\mathbf{r}}_{kk}^{H}\mathbf{w}_k+p_k\tau _p\mathbf{w}_{k}^{H}\tilde{\mathbf{r}}_{kk}\overline{\mathbf{g}}_{kk}^{H}\mathbf{w}_k-\mathbf{w}_{k}^{H}( \overline{\mathbf{G}}_{kk}^{T}\otimes \overline{\mathbf{G}}_{kk} ) \mathbf{w}_k$. However, as for the last three terms in \eqref{LoS_Closed}, it is difficult to directly handle them since $\mathbf{W}_k$ has been divided into many block sub-matrices. To efficiently handle them, we focus on their original expectation form $\mathbb{E} \{ \overline{\mathbf{g}}_{k}^{H}\mathbf{W}_{k}^{H}\overline{\mathbf{g}}_k\overline{\mathbf{g}}_{k}^{H}\mathbf{W}_k\overline{\mathbf{g}}_k \} $. By applying \eqref{vec_eq4}, we have $\mathbb{E} \{ \overline{\mathbf{g}}_{k}^{H}\mathbf{W}_{k}^{H}\overline{\mathbf{g}}_k\overline{\mathbf{g}}_{k}^{H}\mathbf{W}_k\overline{\mathbf{g}}_k \} =\mathbf{w}_{k}^{H}\underset{\mathbf{\Upsilon }_k}{\underbrace{\mathbb{E} \{ ( \overline{\mathbf{g}}_k\overline{\mathbf{g}}_{k}^{H} ) ^T\otimes \overline{\mathbf{g}}_k\overline{\mathbf{g}}_{k}^{H} \} }}\mathbf{w}_k$. Then, we focus on the computation of $\mathbf{\Upsilon }_k\in\mathbb{C} ^{M^2N^2 \times M^2N^2}$, we define $\tilde{\mathbf{G}}_k=\overline{\mathbf{g}}_k\overline{\mathbf{g}}_{k}^{H}\in\mathbb{C} ^{MN \times MN}$ and the $(d,c)$-th element of $\tilde{\mathbf{G}}_k$ is $\tilde{G}_{k,dc}=\overline{g}_{m_1k,n_1}\overline{g}_{m_2k,n_2}^{*}e^{j\theta _{m_1k}}e^{-j\theta _{m_2k}}$, where these variables are defined in Theorem~\ref{Centralized_OBE_closed}. Then, by applying the characteristic of Kronecker-product, we can represent the $(c,d)$-th block sub-matrix of $\mathbf{\Upsilon }_k$ as $\mathbf{\Upsilon }_{k,cd}=\mathbb{E} \{ \tilde{G}_{k,dc}\tilde{\mathbf{G}}_k \} \in \mathbb{C} ^{MN\times MN}$. Finally, by plugging $\tilde{G}_{k,dc}$ into $\mathbf{\Upsilon }_{k,cd}$ and applying the features of random phase-shifts, we can easily compute $\mathbf{\Upsilon }_k$ in closed-form as $\overline{\mathbf{\Upsilon }}_k$. Thus, plugging derived results into \eqref{Centrlized_closed} and applying the property of the generalized Rayleigh quotient, we can easily obtain the closed-form results of $\mathbf{w}_{k}^{*}$ and $\mathrm{SINR}_{k}^{\mathrm{c},\mathrm{UatF},*}$ in Theorem~\ref{Centralized_OBE_closed}.

\vspace*{-0.4cm}
\section{Proof of Corollary~\ref{DG-OBE-Monte}}\label{app_distributed_obe}
\vspace*{-0.2cm}
We first reconstruct the SINR in \eqref{SINR_Distributed}. Note that $\sum_{m=1}^M{\mathbb{E} \{ \mathbf{v}_{mk}^{H}\mathbf{g}_{mk} \}}=\sum_{m=1}^M{\mathbf{w}_{mk}^{H}\mathrm{vec}( \mathbb{E} \{ \mathbf{g}_{mk}\widehat{\mathbf{g}}_{mk}^{H} \} )}$ by applying \eqref{trace} and \eqref{vec_eq3}, where $\mathbf{w}_{mk}=\mathrm{vec}( \mathbf{W}_{mk})\in \mathbb{C} ^{N^2}$. Then, by defining $\mathbf{w}_{k}^{\mathrm{d}}=\left[ \mathbf{w}_{1k}^{T},\mathbf{w}_{2k}^{T},\dots ,\mathbf{w}_{Mk}^{T} \right] ^T\in \mathbb{C} ^{MN^2}$ and $\mathbf{h}_k=[ \mathrm{vec}( \mathbf{g}_{1k}\widehat{\mathbf{g}}_{1k}^{H} ) ^T,\dots ,\mathrm{vec}( \mathbf{g}_{Mk}\widehat{\mathbf{g}}_{Mk}^{H} ) ^T ] ^T\in \mathbb{C} ^{MN^2}$, we can further represent it as $\sum_{m=1}^M{\mathbb{E} \{ \mathbf{v}_{mk}^{H}\mathbf{g}_{mk} \}}=\mathbf{w}_{k}^{\mathrm{d},H}\mathbb{E} \{ \mathbf{h}_k \}$. As for $\mathbb{E} \{ | \sum_{m=1}^M{\mathbf{v}_{mk}^{H}\mathbf{g}_{ml}} |^2 \}$, we have
\setcounter{equation}{45}
\begin{equation}
\begin{aligned}
&\mathbb{E} \{ | \sum\nolimits_{m=1}^M{\mathbf{v}_{mk}^{H}\mathbf{g}_{ml}} |^2 \} \overset{( a )}{=}\mathbb{E} \{ | \sum\nolimits_{m=1}^M{\mathrm{tr}( \mathbf{W}_{mk}^{H}\mathbf{g}_{ml}\widehat{\mathbf{g}}_{mk}^{H} )} |^2 \} \\
&\overset{( b )}{=}\mathbb{E} \{ | \sum\nolimits_{m=1}^M{\mathbf{w}_{mk}^{H}\mathrm{vec}( \mathbf{g}_{ml}\widehat{\mathbf{g}}_{mk}^{H} )} |^2 \} =\mathbb{E} \{ | \mathbf{w}_{k}^{\mathrm{d},H}\mathbf{z}_{kl} |^2 \} \\
&=\mathbf{w}_{k}^{\mathrm{d},H}\mathbb{E} \{ \mathbf{z}_{kl}\mathbf{z}_{kl}^{H} \} \mathbf{w}_{k}^{\mathrm{d}},
\end{aligned}
\end{equation}
where step (a) and step (b) follows from \eqref{trace} and \eqref{vec_eq3}, respectively, with $\mathbf{z}_{kl}=[ \mathrm{vec}( \mathbf{g}_{1l}\widehat{\mathbf{g}}_{1k}^{H} ) ^T,\dots ,\mathrm{vec}( \mathbf{g}_{Ml}\widehat{\mathbf{g}}_{Mk}^{H} ) ^T ] \in \mathbb{C} ^{MN^2}$. Moreover, we have 
\begin{equation}
\begin{aligned}
&\sum\nolimits_{m=1}^M{\mathbb{E} \{ \| \mathbf{v}_{mk} \| ^2 \}}{=}\sum\nolimits_{m=1}^M{\mathrm{tr}( \mathbf{W}_{mk}^{H}\mathbf{W}_{mk}\mathbb{E} \{ \widehat{\mathbf{g}}_{mk}\widehat{\mathbf{g}}_{mk}^{H} \} )}\\
&\overset{( a )}{=}\sum\nolimits_{m=1}^M{\mathbf{w}_{mk}^{H}( \mathbb{E} \{ \widehat{\mathbf{g}}_{mk}\widehat{\mathbf{g}}_{mk}^{H} \} ^T\otimes \mathbf{I}_N )}\mathbf{w}_{mk}=\mathbf{w}_{k}^{H}\mathbf{\Theta }_{k}\mathbf{w}_k,
\end{aligned}
\end{equation}
where step (a) follows from  \eqref{vec_eq4}, with $\mathbf{\Theta }_{k}=\mathrm{diag}\{ ( \mathbb{E} \{ \widehat{\mathbf{g}}_{1k}\widehat{\mathbf{g}}_{1k}^{H} \} ^T\otimes \mathbf{I}_N ) ,\dots ,( \mathbb{E} \{ \widehat{\mathbf{g}}_{Mk}\widehat{\mathbf{g}}_{Mk}^{H} \} ^T\otimes \mathbf{I}_N ) \} \in \mathbb{C} ^{MN^2\times MN^2}$. Combining all above terms, we can rewrite the SINR in \eqref{SINR_Distributed} as
\begin{equation}
\begin{aligned}\label{Dis_SINR_expansion}
&\widetilde{\mathrm{SINR}}_{k}^{\mathrm{LSFD}}=\\
&\frac{p_k| \mathbf{w}_{k}^{\mathrm{d},H}\mathbb{E} \{ \mathbf{h}_k \} |^2}{\mathbf{w}_{k}^{\mathrm{d},H}\!(\! \sum\limits_{l=1}^K{p_l\mathbb{E} \{ \mathbf{z}_{kl}\mathbf{z}_{kl}^{H} \} \!-\!p_k\mathbb{E} \{ \mathbf{h}_k \} \mathbb{E} \{ \mathbf{h}_k \} ^H\!\!+\!\!\sigma ^2\mathbf{\Theta }_{k}} ) \mathbf{w}^{\mathrm{d}}_k}.
\end{aligned}
\end{equation}
Note that $\mathrm{SINR}_{k}^{\mathrm{d}}$ in \eqref{Dis_SINR_expansion} is a generalized Rayleigh quotient with respect to $\mathbf{w}^{\mathrm{d}}_k$. Thus, we can derive the optimal $\mathbf{w}_{k}^{\mathrm{d},*}$ in \eqref{DG_OBE_W_Monte}, which can maximize $\mathrm{SINR}_{k}^{\mathrm{d}}$ with the maximum value $\mathrm{SINR}_{k}^{\mathrm{d},*}=p_k\mathbb{E} \{\mathbf{h}_k\}^H\mathbf{w}_{k}^{\mathrm{d},*}$. Then, by applying the vectorization reversion to $\mathbf{w}_{k}^{\mathrm{d},*}=[ \mathbf{w}_{1k}^{*,T},\dots ,\mathbf{w}_{Mk}^{*,T} ] ^T\in \mathbb{C} ^{MN^2}$ for each AP, we can derive the optimal $\mathbf{W}_{mk}^{*}=\mathrm{vec}^{-1}( \mathbf{w}_{mk}^{*})$.

\section{Proof of Theorem~\ref{Distributed_OBE_closed}}\label{app_distributed_obe_closed}
Based on Theorem~\ref{thm_LSFD_closed}, we can easily compute \eqref{SINR_Distributed} in closed-form as \eqref{SINR_Distributed_Closed} on the top of this page, where $\gamma _{mkl}=\mathrm{tr}( \mathbf{W}_{mk}^{H}\mathbf{B}_{mlk} )$. Then, we have $| \sum_{m=1}^M{\mathrm{tr}\left( \mathbf{W}_{mk}^{H}\overline{\mathbf{R}}_{mk} \right)} |^2=| \sum_{m=1}^M{\mathbf{w}_{mk}^{H}\overline{\mathbf{h}}_{mk}} |^2=| \mathbf{w}_{k}^{\mathrm{d},H}\overline{\mathbf{h}}_k |^2$. Then, by applying results in Appendix~\ref{useful}, we can rewrite $\varphi _{mkl}$, $\upsilon _{mkl}$, and $\gamma _{mkl}$ in \eqref{SINR_Distributed_Closed} as $\varphi _{mkl}=\mathbf{w}_{mk}^{H}( \overline{\mathbf{G}}_{mkk}^{T}\otimes \overline{\mathbf{G}}_{mll} ) \mathbf{w}_{mk}+\mathbf{w}_{mk}^{H}( \overline{\mathbf{G}}_{mkk}^{T}\otimes \check{\mathbf{R}}_{ml} ) \mathbf{w}_{mk}+\mathbf{w}_{mk}^{H}( \widehat{\mathbf{R}}_{mk}^{T}\otimes \overline{\mathbf{G}}_{mll} ) \mathbf{w}_{mk}+\mathbf{w}_{mk}^{H}( \widehat{\mathbf{R}}_{mk}^{T}\otimes \check{\mathbf{R}}_{ml} ) \mathbf{w}_{mk}$, $\upsilon _{mkl}=p_kp_l\tau _{p}^{2}| \mathbf{w}_{mk}^{H}\tilde{\mathbf{r}}_{mlk} |^2$ for $l\in \mathcal{P} _k\backslash \{ k \}$ and $p_kp_l\tau _{p}^{2}| \mathbf{w}_{mk}^{H}\tilde{\mathbf{r}}_{mlk} |+p_{k}^{2}\tau _p\overline{\mathbf{g}}_{mkk}\tilde{\mathbf{r}}_{mkk}^{H}+p_{k}^{2}\tau _p\tilde{\mathbf{r}}_{mkk}\overline{\mathbf{g}}_{mkk}^{H}$ for $l=k$, and $\gamma _{mkl}=\mathbf{w}_{mk}^{H}\mathbf{b}_{mlk}$, respectively. By defining $\mathbf{w}_{k}^{\mathrm{d}}=\left[ \mathbf{w}_{1k}^{T},\mathbf{w}_{2k}^{T},\dots ,\mathbf{w}_{Mk}^{T} \right] ^T\in \mathbb{C} ^{MN^2}$ and based on above results, we can easily construct the diagonal block matrices $\mathbf{\Lambda }_{k}^{( 1 )}$, $\mathbf{\Lambda }_{k}^{( 2 )}$, and $\mathbf{\Lambda }_{k}^{( 4 )}$. Moreover, we have $| \sum_{m=1}^M{\gamma _{mkl}} |^2=| \sum_{m=1}^M{\mathbf{w}_{mk}^{H}\mathbf{b}_{mlk}} |^2=| \mathbf{w}_{k}^{\mathrm{d},H}\mathbf{b}_{lk} |^2$ and $\sum\nolimits_{m=1}^M{| \gamma _{mkl} |^2}=\sum\nolimits_{m=1}^M{| \mathbf{w}_{mk}^{H}\mathbf{b}_{mlk} |^2=\mathbf{w}_{k}^{\mathrm{d},H}\overline{\mathbf{B}}_{lk}\mathbf{w}_{k}^{\mathrm{d}}}$. Thus, we can derive $\mathbf{\Lambda }_{k}^{( 3)}$. Plugging all derived results into \eqref{SINR_Distributed_Closed} and applying the property of the generalized Rayleigh quotient, we can obtain the closed-form results in Theorem~\ref{Distributed_OBE_closed}.

\end{appendices}

\bibliographystyle{IEEEtran}
\bibliography{IEEEabrv,Ref}

\end{document}